\documentclass[a4paper,11pt]{article}
\pdfoutput=1
\usepackage{graphicx}
\usepackage{epsf,amsmath,bbold,amsfonts,stmaryrd}

\usepackage[utf8]{inputenc}
\usepackage{mathrsfs}
\usepackage{appendix}
\usepackage{amssymb}
\usepackage{float}
\usepackage{color}
\usepackage{cite}
\usepackage{hyperref}
\hypersetup{pageanchor=false}
\usepackage{indentfirst}
\usepackage{url}
\usepackage{float}
\usepackage{caption}
\usepackage[numbers,square,comma,sort&compress,merge]{natbib}

\def\mc{\mathcal}

\def\be{\begin{equation}}
\def\ee{\end{equation}}

\def\bea{\begin{eqnarray}}
\def\eea{\end{eqnarray}}

\def\ba{\begin{array}}
\def\ea{\end{array}}

\def\bc{\begin{center}}
\def\ec{\end{center}}

\def\bl{\begin{flushleft}}
\def\el{\end{flushleft}}

\def\br{\begin{flushright}}
\def\er{\end{flushright}}

\def\bi{\begin{itemize}}
\def\ei{\end{itemize}}

\def\bt{\begin{tabular}}
\def\et{\end{tabular}}
\newcommand{\sR}{\mathsf{R}}

\numberwithin{equation}{section}

\begin{document}

\title{\textbf{Einstein-Gauss-Bonnet Black Rings at Large $D$}}
\author{Bin Chen$^{1,2,3}$, Peng-Cheng Li$^{1}$ and Cheng-Yong
Zhang$^{3}$\thanks{bchen01@pku.edu.cn, wlpch@pku.edu.cn, zhangcy0710@pku.edu.cn,
}}

\date{}

\maketitle

\vspace{-10mm}

\begin{center}
{\it
$^1$Department of Physics and State Key Laboratory of Nuclear Physics and Technology,\\Peking University, 5 Yiheyuan Road, Beijing 100871, China\\\vspace{1mm}

$^2$Collaborative Innovation Center of Quantum Matter, 5 Yiheyuan Road, Beijing 100871, China\\\vspace{1mm}

$^3$Center for High Energy Physics, Peking University, 5 Yiheyuan Road, Beijing 100871, China
}
\end{center}

\vspace{8mm}


\vspace{8mm}

\begin{abstract}
We study  the black ring solution in the Einstein-Gauss-Bonnet (EGB) theory at large $D$. By  using the $1/D$ expansion in the near horizon region we derive the effective equations for the slowly rotating black holes in the EGB theory. The effective equations describe the non-linear dynamics of various stationary solutions, including  the EGB black ring, the slowly rotating EGB black hole and the slowly boosted EGB black string. By different embeddings we construct these stationary solutions explicitly.  By performing the perturbation analysis of the effective equations, we obtain the quasinormal modes of the EGB black ring.  We find that thin EGB black ring becomes unstable against non-axisymmetric perturbation.
Furthermore, we numerically evolve the effective equations in a particular case  to study the final state of the instability, and find that  the thin black ring becomes the stable non-uniform
black ring at late time, which gives a relative strong evidence to support the conjecture given in \cite{Tanabe1510}. \end{abstract}
\baselineskip 18pt

\thispagestyle{empty}
\newpage

\section{Introduction}

Black ring is an interesting black object in higher dimensional Einstein gravity. In higher dimensions,
 due to the ineffectiveness  of the uniqueness theorem  the black hole physics are much more rich \cite{Emparan0801,Horowitz2012}. The prototype in higher dimensional black holes is the  five-dimensional black ring, as  the first asymptotically flat black hole  with a non-spherical topology \cite{Emparan0110260}.
Since its discovery, the subsequent developments have greatly enriched our understanding of the black hole  physics in five dimensions. Nevertheless, the
black hole solutions in more than five dimensions has  not been fully explored. For example, unlike the case in five dimensions, the exact  black ring solution in
higher dimensions is still lacking. All the  constructions of the black rings in higher dimensions were based on the  numerical methods \cite{Dias1510} and analytical
approximation method such as the blackfold method \cite{Emparan0708, Armas1402}.

An important aspect of the black ring is its dynamical stability. First of all, the thin black ring suffers from the Gregory-Laflamme (GL)-like instability \cite{Hovdebo0601079, Elvang0608076}. This is due to the fact that the thin black ring is equivalent to a critically boosted black string, which suffers from the Gregory-Laflamme (GL) instability \cite{Gregory9301052, Harmark0701002}. Moreover,  the fat black ring is unstable under the radial perturbations \cite{Elvang0608076}. These instabilities have been verified by the  numerical study in five dimensions \cite{Figueras1107, Santos1503,  Figueras1512}. In particular,  it was found in \cite{Figueras1512} that for a sufficiently thin ring the GL mode is dominant, and
its  non-linear evolution leads to the pinch-off of the ring, indicating thus the violation of the weak cosmic censorship conjecture. However,
 for a certain thin ring, a new elastic-type instability was identified and it becomes dominant during the evolution of the black ring, leading the system to  a spherical black hole.

It has recently been shown that in the limit of large $D$ dimensions the
gravitational field of a black hole is confined in the near horizon region \cite{Emparan1302}, thus the black hole can be viewed as a surface or a thin membrane embedded in a background geometry \cite{Emparan1406,  Emparan1504, Bhattacharyya1504, Suzuki1505, Bhattacharyya1511, Dandekar1607, Bhattacharyya1704,Bhattacharyya1805}. The membrane is described by the  way it is embedded in the background and its dynamics is governed by the effective equations obtained from the $1/D$ expansion of the Einstein equations.
Solving the effective equations with different embeddings, one can construct the black hole solutions with various topologies, and furthermore to study their dynamics \cite{Suzuki1506, Emparan1506, Tanabe1510,  Tanabe1511, Emparan1602, Sadhu1604,  Herzog1605,
Tanabe16050811, Tanabe16050885, Rozali1607,  Chen1702, Rozali1707,Chen1804}. For example, in \cite{Tanabe1510} by solving the effective equations of slowing rotating black holes  with the embedding into a flat background in the ring coordinates, the black ring solution was constructed analytically in the $1/D$ expansion. Then, by performing the perturbation analysis of the effective equations,  the thin black ring was found to suffer from the GL-like instability, which is consistent with previous analyses and numerical studies.
By performing the $1/\sqrt{D}$ instead of $1/D$ expansion of the Einstein equations and considering higher order corrections of the effective equations,  the elastic instability of the black ring was investigated in \cite{Tanabe16050811}. It is interesting to note that from a broader perspective, the large $D$ expansion method of the black hole is similar to the blackfold method \cite{Emparan0902, Emparan0910}.
The difference is that the spacetime dimension $D$ provides a natural expansion parameter, independent of the specific solution.

The purpose of this work is to study the black ring solutions in the Einstein-Gauss-Bonnet (EGB) gravity by using the large $D$
expansion method. In the dimensions higher than four, the Lovelock higher-curvature gravity of various orders is the natural generalization of the Einstein gravity. One attractive feature of the Lovelock gravity is that its equations of motion are
still the second order differential equations such that the fluctuations around the vacuum do not have ghost-like mode.  Among all the Lovelock gravities, the second-order Lovelock gravity, the EGB gravity, is of particular interest. It includes the quadratic terms of the curvature tensors which appear as the leading-order correction in the low energy effective action of the heterotic string theory \cite{Zwiebach1985, Boulware1985}. Although the generalization of the spherically symmetric Schwarzschild solution in the EGB gravity has been known for quite a long time \cite{ Boulware1985, Wheeler1986}, the exact solution with non-spherical topology is still unknown. There are many efforts on the black string solutions in the EGB gravity based on the numerical analysis in $D\geq5$ \cite{Kobayashi2005,Suranyi2009, Brihaye2010}. However, for the black ring solutions in the EGB gravity, numerical discussion has been implemented only for $D=5$ \cite{Kleihaus0912}, with the  higher dimensional cases being  basically unexplored.
 The large $D$ expansion method developed recently offers a promising framework to address this issue.

 The large $D$ study of the EGB black holes was initiated in \cite{Chen1511}, in which the quasinormal modes in the large $D$ limit
have been computed. It has been found that similar to the case of the Einstein gravity the quasinormal modes can be split into two classes, non-decoupling ones and decoupled ones \cite{Emparan1406}. This suggests that the large $D$ EGB black holes can also be viewed as a membrane embedded in a background, so a large $D$ effective theory should be expected. Indeed, the large $D$ effective theory of the EGB black holes has been developed and the instability of the static spherical solution has been investigated in \cite{Chen1703}.  In \cite{Chen1707} by using the large $D$ effective theory the uniform black string was obtained analytically, the non-uniform black string was obtained numerically, and the instability was  investigated. Like the case in the Einstein gravity \cite{Emparan1506},
the thin EGB black strings are unstable to developing inhomogeneities along the strings, and at late
time they settle to the stable non-uniform black strings.

Intuitively, one can imagine a black ring as a rotating bent string such that the
centrifugal force balances the string tension. According to the analysis of the blackfold
method \cite{Emparan0708}, a thin black ring can be identified to be a boosted black string and its horizon
angular velocity is determined to be
\be
\Omega_H=\frac{1}{\sqrt{D-3}}\frac{1}{R},\label{condition}
\ee
where $R$ is the ring radius.  At large $D$ since the string tension is small compared with its mass \cite{Emparan1302}, from the balance condition the horizon angular velocity of the black ring should be small as well. Thus the condition (\ref{condition})  is  universal,  not just for the thin black rings. In the following the black holes with the horizon angular velocity of order $\mc O(1/\sqrt{D})$ will be referred to as the slowly rotating black holes. On the other hand, from \cite{Chen1707} one can
find that  the tension of a EGB black string is small compared with its mass at large $D$.
 The effect of the GB term is contained in the $1/D$ corrections of the mass and the
tension. Therefore, it should be possible to construct a EGB black ring solution by bending and rotating the EGB black string, with the horizon angular velocity being of $\mc O(1/\sqrt{D})$.

In this paper, we study the EGB
black ring and  its dynamical instabilities by applying the large $D$ effective theory.
 In section \ref{section2} we solve the EGB equations with proper metric ansatz and derive the effective equations for the slowly rotating black holes at large $D$. We
obtain  stationary solutions from the effective equations and discuss thermodynamic quantities of the solutions. In section \ref{section3}
we construct the EGB black ring solution explicitly by solving the effective equations with the embedding of the leading order solution into flat background geometry in the ring coordinates. In section \ref{section4} we investigate the stability of the EGB black ring solution by perturbatively analyzing the effective equations. We find that the thin black ring suffers from the GL-like instability, and then we clarify the effect of the GB term on the instability. Furthermore, we numerically study the non-linear evolution of the GL-like instability  to find the end point of the instability. In section \ref{section5} we have parallel discussions on the slowly rotating black hole and the slowly boosted black string. We show that in the large ring radius limit, the EGB black ring is equivalent to the boosted EGB
black string when the boost velocity takes a specific value. We end with a summary and some discussions in section \ref{section6}.

\section{Effective equations}\label{section2}
In this section we study the large $D$ effective theory for the slowly rotating black holes in the EGB gravity. By solving
the EGB equations, we derive the effective equations, which  depends on the mass and the momentum density of a dynamical black object. Then we study the general properties of the stationary solutions of the effective equations without specifying the embeddings of the solutions. In the following we use $1/n$ as the expansion parameter instead of $1/D$, where
\be
n=D-4.
\ee
\subsection{Set up}
We consider the $D$-dimensional Einstein-Hilbert action supplemented by the GB term:
\be
I=\frac{1}{16\pi G}\int d^Dx\sqrt{-g}\biggl(R_g+\alpha L_{GB}\biggl),
\ee
with
\be
L_{GB}=R_{\mu\nu\lambda\delta}R^{\mu\nu\lambda\delta}-4R_{\mu\nu}R^{\mu\nu}+R_g^2,
\ee
where $\alpha$ is the GB coefficient. From the action, we obtain the equations of motion for the metric
\be\label{EGBequations}
R_{\mu\nu}-\frac{1}{2}g_{\mu\nu}R_g+\alpha H_{\mu\nu}=0,
\ee
where
\be
H_{\mu\nu}=-\frac{1}{2}g_{\mu\nu}L_{GB}+2(R_gR_{\mu\nu}-2R_{\mu\gamma}R^{\gamma}_{\,\,\nu}+2R^{\gamma\delta}R_{\gamma\mu\nu\delta}
+R_{\mu\gamma\delta\lambda}R_{\nu}^{\,\,\gamma\delta\lambda}).
\ee
We make the following metric ansatz for the slowly
rotating black holes in the ingoing Eddington-Finkelsteins coordinates
\be\label{metricansatz}
ds^2=-Adv^2+2(u_v dv+u_a dX^a)dr+r^2G_{ab} dX^adX^b-2C_a dv dX^a+r^2 H^2 d\Omega_n^2.
\ee
Here $X^a$ denote the two coordinates $(z, \Phi)$ in which  $\Phi$ can be specified to be the rotational direction. By gauge choice we may set $u_a dX^a dr=0$. We consider the case that all components of the metric are functions of $(v, r, X^a)$.

  It has been known that up to the linear order of the rotation parameter $a$ the higher dimensional slowly rotating black holes in the EGB theory are easy to construct, since in this case $a$ only appears in the component $g_{v\Phi}$ which can be treated as a perturbation \cite{Kim0711}. Now we are considering the slowly rotating black holes in the EGB theory with a small $a=\mc O(1/\sqrt{n})$, the form of the metric should be the same as that in the Einstein gravity \cite{Chen1702}. Note that in the above metric ansatz the information of the horizon shape  is
undetermined, so that we can discuss various stationary solutions including the black ring, black hole and  the
(boosted) black string from the same ansatz.

  In order to do the $1/n$ expansion properly we need to specify the large $D$ behaviors of the metric functions. According to the previous analysis,  the EGB black rings are among the slowly rotating black holes. We assume that the horizon velocity is of  $\mc O(1/\sqrt{n})$, so we have
  \be
  C_\Phi=\mc O(1/\sqrt{n}).
  \ee
Define
\be\label{phiandPhi}
\phi\equiv\sqrt{n}\Phi,
\ee
 in terms of which
 \be
 g_{v\phi}=\mc O(1/n).
 \ee
Then the metric ansatz can be  written as
\be
ds^2=-Adv^2+2u_vdvdr+r^2 G_{ab}dx^adx^b-2C_advdx^a+r^2H^2 d\Omega^2_n,
\ee
where $x^a=(z,\phi)$. At large $D$ the radial gradient becomes dominant, i.e. $\partial_r= \mc O(n)$, $\partial_v= \mc O(1)$, $\partial_a =\mc O(1)$, so in the near horizon region of the black hole it is better to use a new radial coordinate $\sR$ defined by
\be
\sR=\Big(\frac{r}{r_0}\Big)^n,
\ee
such that $\partial_\sR= \mc O(1)$, where $r_0$ is a horizon length scale which can be set to unity $r_0=1$. The large $D$ scalings of the metric functions are respectively
\be
A= \mc O(1), \quad u_v=\mc O(1),\quad C_a=\mc O(n^{-1}), \quad G_{zz}=1+\mc O(n^{-1}),
\ee
\be
G_{z\phi}=\mc O(n^{-2}),\quad G_{\phi\phi}=\frac{G(z)^{2}}{n}\Big(1+\frac{G_{\phi\phi}^{(0)}}{n}+\mc O(n^{-2})\Big), \quad H=H(z).
\ee
At the asymptotic infinity ( $\sR\to\infty$) both $\partial_v$ and $\partial_\phi$ are expected to be the Killing vectors, which implies that the metric functions at the asymptotic infinity are
independent of $v$ and $\phi$. Thus the function $H$ and the leading order of $G_{\phi\phi}$ depend only on $z$. Other metric components are still the functions of $(v, \sR, x^a)$.

On the other hand, to solve the EGB equations we need to specify the boundary conditions at large $R$. They are given by \cite{Emparan1406}
\be
A=1+\mc O(\sR^{-1}),\quad C_a=\mc (\sR^{-1}), \quad G_{z\phi}=\mc (\sR^{-1}).
\ee
Besides, the solutions have to be regular at the horizon as well.

In the following as in the case of the EGB black string \cite{Chen1707} we use  $\tilde{\alpha}$ instead of $\alpha$ in doing the $1/n$ expansion, with
\be
\tilde{\alpha}=\alpha n(n+1).
\ee
In some of the literatures, it was exactly  $\tilde{\alpha}$ that is called the GB coefficient. In this work we focus on the case $\tilde{\alpha}=\mc O(1)$ and the value of $\tilde{\alpha}$ is not restricted to be positive, despite the fact that it is positive definite in the heterotic string theory \cite{Boulware1985}.

\subsection{Effective equations}
At the leading order of the $1/n$ expansion, the EGB equations (\ref{EGBequations}) only contain $\sR$-derivatives
so they can be solved by performing $\sR$-integrations.  The leading order solutions are easily obtained as
\be\label{leadingordersol}
A=1+\frac{u_v^2}{2\tilde{\alpha}}\Big( 1-\Sigma \Big),\qquad u_v=\frac{H(z)}{\sqrt{1-H'(z)^2}},
\ee
\be
C_a=\frac{1}{n}\frac{u_v^2 p_a}{2\tilde{\alpha}p_v}\Big( -1+\Sigma\ \Big),\qquad G_{zz}=1+\mc O(n^{-2}),
\ee
\bea
G_{z\phi}&=&\frac{1}{n^2}\Bigg[\frac{u_v^2p_zp_\phi}{2\tilde{\alpha}p_v^2}\Big(\Sigma-1\Big)+\Bigg(\ln \frac{1+\Sigma}{2}+\frac{\pi}{4}-\arctan \Sigma-\frac{u_v^2}{2(u_v^2+2\tilde{\alpha})}\ln\frac{1+\Sigma^2}{2}
\Bigg)\times\nonumber\\
&&\frac{u_v^3(G(p_z\partial_\phi p_v+p_\phi\partial_z p_v)+p_v(2p_\phi G'-G(\partial_\phi p_z+\partial_zp_\phi)))}{Gp_v^2(\tilde{\alpha}+u_v^2)}\Bigg],
\eea
\bea
G_{\phi\phi}^{(0)}&=&-\Big(2+\frac{2u_v^2 G'H'}{GH}\Big)\ln\sR+\frac{u_v^2p_\phi^2}{2\tilde{\alpha}G^2p_v^2}\Big(\Sigma-1\Big)\nonumber\\
&&-\Bigg(\ln\frac{1+\Sigma}{2}+\frac{\pi}{4}-\arctan \Sigma-\frac{u_v^2}{2(u_v^2+2\tilde{\alpha})}\ln\frac{1+\Sigma^2}{2}\Bigg)\times\nonumber\\
&&\,\frac{2u_v^2(G^2Hp_v^2+u_v^2Gp_v^2G'H'+u_vH(p_v \partial_\phi p_\phi -p_\phi\partial_\phi p_v))}{HG^2p_v^2(\tilde{\alpha}+u_v^2)},
\eea
where
\be\label{Sigmavzphi}
\Sigma=\sqrt{1+\frac{4\tilde{\alpha} p_v}{\sR u_v^2}}.
\ee
In the above expressions, $p_v(v,z)$ and $p_a(v,z)$ are the integration functions of $\sR$-integrations of the EGB equations. Physically they can be viewed as the mass and the momentum density of the solution. The horizon of this dynamical black hole solution is at
\be\label{horizon}
\sR_H=\frac{u_v^2 }{u_v^2+\tilde{\alpha}}p_v(v,x^a).
\ee
Note that the functions $G(z)$ and $H(z)$ are undetermined. As we will see in the following they actually encode the information on the topology of the event
horizon and can be determined by embedding the leading order solution into a specific
background. Different embeddings determine different forms
of the functions $G(z)$ and $H(z)$, and thus different black objects,  the black ring,
the black hole or the black string.

By performing the $``D-1+1"$ decomposition on a $r=$ {\it constant} surface, the momentum constraint tells us that the function $H(z)$ should satisfy the following relation
\be
1-H'(z)^2+H(z)H''(z)=0.
\ee
In this case $u_v$ can be treated as a constant, so we have
\be\label{kappahat}
\frac{\sqrt{1-H'(z)^2}}{H(z)}=2\hat{\kappa},
\ee
where $\hat{\kappa}$ is a constant and is related to the surface gravity of the horizon.
Solving the next-to-leading order  EGB equations in the $1/n$ expansion we obtain three equations for $p_v(v,x^a)$ and $p_a(v,x^a)$. These equations are called the effective equations for the slowly rotating EGB black holes. These equations are
\be\label{Effeq1}
\partial_v p_v+\frac{H'}{H}p_z-\frac{H'}{2\hat{\kappa}H}\partial_z p_v-\frac{\partial_\phi^2p_v}{2\hat{\kappa}G^2}+\frac{\partial_\phi p_v}{G^2}=0,
\ee
\bea\label{Effeq2}
&&\partial_v p_z-\frac{1+8\tilde{\alpha}\hat{\kappa}^2+32\tilde{\alpha}^2\hat{\kappa}^4}{2\hat{\kappa}(1+4\tilde{\alpha}\hat{\kappa}^2)(1+8\tilde{\alpha}\hat{\kappa}^2)G^2}\partial_\phi^2 p_z
\nonumber\\
&&+\frac{2\tilde{\alpha}\hat{\kappa}}{(1+4\tilde{\alpha}\hat{\kappa}^2)(1+8\tilde{\alpha}\hat{\kappa}^2)G^2}\Bigg(-\partial_\phi\Big(\frac{p_z\partial_\phi p_v}{p_v}\Big)+\partial_z\partial_\phi p_\phi-\partial_\phi\Big(\frac{ p_\phi \partial_z p_v}{ p_v}\Big)\Bigg)\nonumber\\
&&-\frac{(1+4\tilde{\alpha}\hat{\kappa}^2+32\tilde{\alpha}^2\hat{\kappa}^4)H'}{2\hat{\kappa}(1+4\tilde{\alpha}\hat{\kappa}^2)(1+8\tilde{\alpha}\hat{\kappa}^2)H}\partial_z p_z
+\frac{p_\phi}{G^2 p_v}\partial_\phi p_z+\Big(\frac{p_z}
{G^2p_v}+\frac{(1+4\tilde{\alpha}\hat{\kappa}^2+32\tilde{\alpha}^2\hat{\kappa}^4)G'}
{\hat{\kappa}(1+4\tilde{\alpha}\hat{\kappa}^2)(1+8\tilde{\alpha}\hat{\kappa}^2)G^3}\Big)\partial_\phi p_\phi\nonumber\\
&&+\Big(1-\frac{4\tilde{\alpha}\hat{\kappa} p_z H'}{(1+4\tilde{\alpha}\hat{\kappa}^2)(1+8\tilde{\alpha}\hat{\kappa}^2)Hp_v}\Big)\partial_z p_v-\frac{p_\phi((1+12\tilde{\alpha}\hat{\kappa}^2+32\tilde{\alpha}^2\hat{\kappa}^4)Gp_z-4\tilde{\alpha}\hat{\kappa} m G')}{(1+4\tilde{\alpha}\hat{\kappa}^2)(1+8\tilde{\alpha}\hat{\kappa}^2)G^3p_v^2}\partial_\phi p_v\nonumber\\
&&
+\frac{H'}{Hp_v}p_z^2-\frac{G'}{G^3p_v}p_\phi^2+\frac{1+4\tilde{\alpha}\hat{\kappa}^2+32\tilde{\alpha}^2\hat{\kappa}^4-8(\kappa^2+8\tilde{\alpha}\hat{\kappa}^2+32\tilde{\alpha}^2\hat{\kappa}^6)H^2}{2\hat{\kappa}(1+4\tilde{\alpha}\hat{\kappa}^2)(1+8\tilde{\alpha}\hat{\kappa}^2)H^2}p_z\nonumber\\
&&+\frac{(1+4\tilde{\alpha}\hat{\kappa}^2+32\tilde{\alpha}^2\hat{\kappa}^4)p_v(HG'^2H'+G(G'-HH'G''))}{4\hat{\kappa}^2(1+4\tilde{\alpha}\hat{\kappa}^2)(1+8\tilde{\alpha}\hat{\kappa}^2)G^2H^2}=0,
\eea

\bea\label{Effeq3}
&&\partial_v p_\phi-\frac{1+4\tilde{\alpha}\hat{\kappa}^2+32\tilde{\alpha}^2\hat{\kappa}^4}{2\hat{\kappa}(1+4\tilde{\alpha}\hat{\kappa}^2)(1+8\tilde{\alpha}\hat{\kappa}^2)G^2}\partial_\phi^2 p_\phi
-\frac{4\tilde{\alpha}\hat{\kappa} p_\phi}{(1+4\tilde{\alpha}\hat{\kappa}^2)(1+8\tilde{\alpha}\hat{\kappa}^2)G^2p_v}\partial_\phi^2 p_v\nonumber\\
&&+\frac{4\tilde{\alpha}\hat{\kappa} p_\phi}{(1+4\tilde{\alpha}\hat{\kappa}^2)(1+8\tilde{\alpha}\hat{\kappa}^2)G^2p_v^2}(\partial_\phi p_v)^2
-\frac{4\tilde{\alpha}\hat{\kappa}}{(1+4\tilde{\alpha}\hat{\kappa}^2)(1+8\tilde{\alpha}\hat{\kappa}^2)G^2p_v}\partial_\phi p_\phi \partial_\phi p_v\nonumber\\
&&+\frac{2\tilde{\alpha}\hat{\kappa} H'}{(1+12\tilde{\alpha}\hat{\kappa}^2+32\tilde{\alpha}^2\hat{\kappa}^4)H}\partial_\phi p_z
-\frac{(1+8\tilde{\alpha}\hat{\kappa}^2+32\tilde{\alpha}^2\hat{\kappa}^4)H'}{2\kappa(1+4\tilde{\alpha}\hat{\kappa}^2)(1+8\tilde{\alpha}\hat{\kappa}^2)H}\partial_z p_\phi
+\frac{2p_\phi}{G^2 p_v}\partial_\phi p_\phi\nonumber\\
&&-\Big(\frac{p_\phi^2}{G^2p_v^2} +\frac{(1+8\tilde{\alpha}\hat{\kappa}^2+32\tilde{\alpha}^2\hat{\kappa}^4)Gp_v^2G'H'
+2\hat{\kappa}^2G^2p_v((1+4\tilde{\alpha}\hat{\kappa}^2+32\tilde{\alpha}^2\hat{\kappa}^2)Hp_v+2\tilde{\alpha}\hat{\kappa} p_zH')}
{2\hat{\kappa}^2(1+4\tilde{\alpha}\hat{\kappa}^2)(1+8\tilde{\alpha}\hat{\kappa}^2)G^2Hp_v^2}\Big)\partial_\phi p_v\nonumber\\
&&-\frac{2\tilde{\alpha}\hat{\kappa} p_\phi H'}{(1+4\tilde{\alpha}\hat{\kappa}^2)(1+8\tilde{\alpha}\hat{\kappa}^2)Hp_v}\partial_zp_v +\Big(\frac{H'p_z}{Hp_v}+\frac{(1+8\tilde{\alpha}\hat{\kappa}^2+32\tilde{\alpha}^2\hat{\kappa}^4)G'H'}{\hat{\kappa}(1+4\tilde{\alpha}\hat{\kappa}^2)(1+8\tilde{\alpha}\hat{\kappa}^2)GH }\Big)p_\phi=0.
\eea
From these equations we can obtain the
stationary solutions and study the non-linear dynamics of the dynamical black holes. These equations depend on the forms of the functions $G(z)$ and $H(z)$. In order to  solve these equations, we have to find the appropriate embeddings and determine  the forms of $G(z)$ and $H(z)$.

\subsection{Stationary solutions}
To obtain the stationary solutions we need to assume the existence of two Killing vectors $\mathsf{K}_v=\partial_v$ and  $\mathsf{K}_\phi=\partial_\phi$,
then we have
\be
p_v=p_v(z), \quad p_a=p_a(z).
\ee
From (\ref{Effeq1}) we get
\be\label{staionarysol:pv}
p_z(z)=\frac{p_v'(z)}{2\hat{\kappa}}.
\ee
Plugging this into (\ref{Effeq3}) we obtain
\be\label{staionarysol:pphi}
p_\phi(z)=\hat{\Omega}_H G(z)^2 p_v(z).
\ee
Here $\hat{\Omega}_H$ is introduced as an integration constant independent of $z$. By setting $p_v(z)=e^{P(z)}$ and substituting the solutions (\ref{staionarysol:pv}) and (\ref{staionarysol:pphi}) into (\ref{Effeq2}), we obtain an equation for $P(z)$,
\bea\label{stationarysol:Py}
&&P''(z)-\frac{H'}{H}P'(z)\nonumber\\
&&-\Bigg[\frac{G'^2}{G^2}+\frac{G'}{GHH'}-\frac{G''}{G}-\hat{\Omega}_H^2\frac{GG'(1-H'^2)}{HH'}
\frac{H^4+\tilde{\alpha}H^2(1-H'^2)+3\tilde{\alpha}^2(1-H'^2)^2}{H^4+\tilde{\alpha}H^2(1-H'^2)+2\tilde{\alpha}^2(1-H'^2)^2}\Bigg]=0.
\nonumber\\
\eea
To solve this equation we need to specify the functions $G(z)$ and $H(z)$, which can be implemented by embedding the solutions into a specific background.

The physical meaning of $\hat{\Omega}_H$
will become manifest from the metric. The $(v,\Phi)$ part of the leading order metric of the above stationary solutions is written as
\be
ds^2_{(v,\Phi)}=-Adv^2+G(z)^2\Bigg(d\Phi-\frac{\hat{\Omega}_H}{\sqrt{n}}(1-A)dv\Bigg)^2,
\ee
where we have used the relation $C_\phi=\frac{1}{n}\frac{p_\phi}{p_v}(1-A)$, with $A$ given by (\ref{leadingordersol}). It is easy to see that the horizon angular velocity is related to $\hat{\Omega}_H$ by
\be
\Omega_H=\frac{\hat{\Omega}_H}{\sqrt{n}}.
\ee
 Then we see that the event horizon $\sR=\sR_H$ is the Killing horizon of a Killing vector
\be
\xi=\partial_v+\hat{\Omega}_H\partial_\phi.
\ee
Due to the existence of the Killing vector $\xi$, the associated surface gravity is given by
\bea\label{surfacegra}
\kappa&=&-\frac{\partial_r(\xi_\mu \xi^\mu)}{2\xi_r}\Bigg|_{\sR=\sR_H}\nonumber\\
&=&\frac{n(1+4\tilde{\alpha}\hat{\kappa}^2)\hat{\kappa}}{1+8\tilde{\alpha}\hat{\kappa}^2}.
\eea
As we mentioned before, $\hat{\kappa}$ is a constant and so is the surface gravity $\kappa$. This is expected for the stationary solutions.

From the above general stationary solutions we can read the corresponding thermodynamic quantities by using the Komar integral  and the Wald formula \cite{Wald1993}. At first let us write down the leading order metric
\bea
ds^2&=&-\Big(1+\frac{1-\Sigma }{8\tilde{\alpha}\hat{\kappa}^2} \Big)dv^2+\frac{dvdr}{\hat{\kappa}}+r^2G(z)^2d\Phi^2+r^2 dz^2\nonumber\\
&&+\Omega_H G(z)^2\frac{1-\Sigma}{4\tilde{\alpha}\hat{\kappa}^2}dvd\Phi+r^2H^2d\Omega_n^2,
\eea
where the $\mc O(1/n)$ terms are omitted. The mass and the angular momentum are given respectively by
\bea
\mathcal{M}&=&\frac{n\hat{\kappa}\Omega_n}{4 G_D}\int dz\, G(z) H(z)^n p_v(z),\label{stationarysol:mass}\\
\mathcal{J}&=&\frac{n\hat{\kappa}\Omega_n}{4 G_D}\int dz\, \Omega_H G(z)^3 H(z)^n p_v(z).\label{stationarysol:angularmo}
\eea
From (\ref{surfacegra}) one finds the temperature is
\be\label{stationarysol:tem}
T_H=\frac{\kappa}{2\pi}=\frac{n}{2\pi}\frac{(1+4\tilde{\alpha}\hat{\kappa}^2)\hat{\kappa}}{1+8\tilde{\alpha}\hat{\kappa}^2}.
\ee
The entropy of a black object in the EGB theory can be written as an integral over the event horizon
via the Wald formula
\be
S=\frac{1}{4G_D}\int_{\Sigma_h}d^{n+2}x\sqrt{h}\, (1+2\alpha \tilde{R}),
\ee
where $h$ is the determinant of the induced metric on the event horizon and $\tilde{R}$ is the event horizon curvature. For the stationary solutions
\be
\sqrt{h}=\frac{p_v(z)}{1+4\tilde{\alpha}\hat{\kappa}^2}G(z)H(z)^n,\qquad \tilde{R} =1+8\tilde{\alpha}\hat{\kappa}^2,
\ee
thus
\be\label{stationarysol:entropy}
S=\frac{\Omega_n \pi(1+8\tilde{\alpha}\hat{\kappa}^2)}{2G_D(1+4\tilde{\alpha}\hat{\kappa}^2)}\int dz\, G(z) H(z)^n p_v(z).
\ee
From the above results we see that the thermodynamic quantities of the slowly rotating black holes in the EGB gravity satisfy the Smarr formula at the leading order
of the $1/n$ expansion,
\be
\mathcal{M}=T_H S+\Omega_H \mathcal{J}.
\ee
Note that since the horizon angular velocity $\Omega_H$ is of $\mc O(1/\sqrt{n})$, $\Omega_H \mathcal{J}$ is of $\mc O(1)$. Comparing with other terms in the Smarr formula, we find that $\Omega_H \mathcal{J}$ does not contribute to the Smarr formula at the leading order of the $1/n$ expansion.

\section{EGB black ring}\label{section3}
In this section we study the EGB black ring at large $D$. By embedding the leading order metric into the flat background in the ring coordinates, we determine the functions $G(z)$ and $H(z)$ and  then analytically obtain the EGB black ring solution.
\subsection{EGB black ring solution}
The $D=n+4$ dimensional flat metric in the ring coordinates  is of the form
 \be\label{ringcoordinate}
 ds^2=-dt^2+\frac{R^2}{(R+r \cos\theta)^2}\Biggl[\frac{R^2dr^2}{R^2-r^2}+(R^2-r^2)d\Phi^2+r^2(d\theta^2+\sin^2\theta d\Omega_n^2)\Biggl],
 \ee
 where $R$ is the ring radius, $0\leq r\leq R$, $0\leq\theta\leq\pi$ and $0\leq \Phi\leq2\pi$. $r=0$ is the origin of the ring coordinate.
 Obviously, the topology of a $r=$ {\it constant} surface is $S^1\times S^{n+1}$.
 On the other hand, in the far region of the black hole where $\sR\gg1$, the leading order metric we obtained before behaves like
  \be\label{leadordermetric}
 ds^2\simeq -dv^2+\frac{2H}{\sqrt{1-H'^2}}dvdr+r^2dz^2+r^2G^2d\Phi^2+r^2 H^2 d\Omega^2_n.
 \ee
On a $r=$ {\it constant} surface  the induced metric in the far region  and the metric from the near horizon region solution in the large $D$ limit should  match. Considering the surface $r=r_0=1$, which is in the far region since  $\sR\gg1$ still corresponds to $r=r_0+\mc O(1/n)$, and comparing (\ref{ringcoordinate}) with (\ref{leadordermetric}),  we obtain the following identifications
  \be\label{blackring:identifications}
 H=\frac{R\sin\theta}{R+\cos\theta},\qquad G=\frac{R\sqrt{R^2-1}}{R+\mathrm{cos}\theta},\qquad \frac{d\theta}{dz}=\frac{R+\mathrm{cos}\theta}{R}.
 \ee
Hence $G(z)$ and $H(z)$ are determined. From (\ref{kappahat}) it is easy to find that
\be\label{blakcring:kappa}
\hat{\kappa}=\frac{\sqrt{R^2-1}}{2R}.
\ee
 For the EGB black ring solution the associated surface gravity is still given by (\ref{surfacegra}). By these identifications we can obtain the leading order EGB black ring solution
 as long as we solve the effective equations. Note that since $R\geq r$, we find $R\geq 1$.  The case that $R\gg1$ corresponds to the thin black ring, while the case that  $R\simeq1$ corresponds to the fat black ring.

To solve the effective equations it is convenient to introduce the variable
\be
y=\cos\,\theta.
\ee
 For the stationary solution with $\partial_v$ and $\partial_\phi$ being the Killing vectors,  we have
 \be
 p_v=e^{P(y)},\qquad p_z=p_z(y),\qquad p_\phi=p_\phi(y).
 \ee
 From (\ref{staionarysol:pv}) we find
 \be
 p_z(y)=-\frac{(R+y)\sqrt{1-y^2}}{\sqrt{R^2-1}}p_v'(y).
 \ee
 Then from (\ref{staionarysol:pphi}) we obtain
 \be
 p_\phi(y)=\hat{\Omega}_H\frac{R^2(R^2-1)}{(R+y)^2}p_v(y).
 \ee
 Furthermore from (\ref{stationarysol:Py}) we get the equation for $P(y)$
 \bea\label{Py:blackring}
 &&P''(y)+\frac{2}{R+y}P'(y)-\frac{R}{(R+y)(1+R y)}\nonumber\\
 &&\qquad+\,\hat{\Omega}_H^2\frac{R^2(R^2-1)^2}{(R+y)^4(1+Ry)}
 \frac{(R^2+2\tilde{\alpha}(R^2-1))(R^2+\tilde{\alpha}(R^2-1))}{R^4+\tilde{\alpha} R^2(R^2-1)+2\tilde{\alpha}^2(R^2-1)^2}=0.
 \eea
 The above equation contains a pole at $y=-1/R$, which leads to a logarithmic divergence at $y=-1/R$ for the solution $P(y)$. In order to have
 a regular solution we need $\hat{\Omega}_H$ to be
 \be\label{angularvelocity:blackring}
 \hat{\Omega}_H=\frac{\sqrt{R^2-1}}{R^2}\sqrt{\frac{R^4+\tilde{\alpha} R^2(R^2-1)+2\tilde{\alpha}^2(R^2-1)^2}{(R^2+2\tilde{\alpha}(R^2-1))(R^2+\tilde{\alpha}(R^2-1))}}.
 \ee
This is similar to the black rings in five \cite{Emparan0110260}  or higher dimensions \cite{Emparan0708}: the horizon angular velocity is determined by the regularity condition or the dynamical balance condition.
In the above discussion, we allow the GB coefficient  $\tilde{\alpha}$ to take arbitrary value. However, in order to have a regular solution, from (\ref{angularvelocity:blackring}) we need
\be
R^2+2\tilde{\alpha}(R^2-1)>0.
\ee
As $R^2\geq 1$, we get the constraint on $\tilde{\alpha}$ in order to have a regular black ring solution:
\be \label{GBconstraint}
\tilde{\alpha} > -\frac{1}{2}.
\ee
From (\ref{surfacegra}) and (\ref{blakcring:kappa}) we can see that this is consistent with the requirement that the surface gravity is physical. In fact, from the study on the static EGB string \cite{Chen1707} and the static EGB black holes \cite{Chen1703}, we found that the surface gravities for these black objects always satisfied
\be
\kappa =\frac{C}{1+2\tilde{\alpha}},
\ee
where $C$ is a positive constant. In order to have a physical surface gravity, we need (\ref{GBconstraint}).  Thus, a remarkable result from our study is that the black objects in the EGB theory only exist when the GB coefficient $\tilde{\alpha}$ obeys the constraint (\ref{GBconstraint}).
  \begin{figure}[t]
 \begin{center}
  \includegraphics[width=65mm,angle=0]{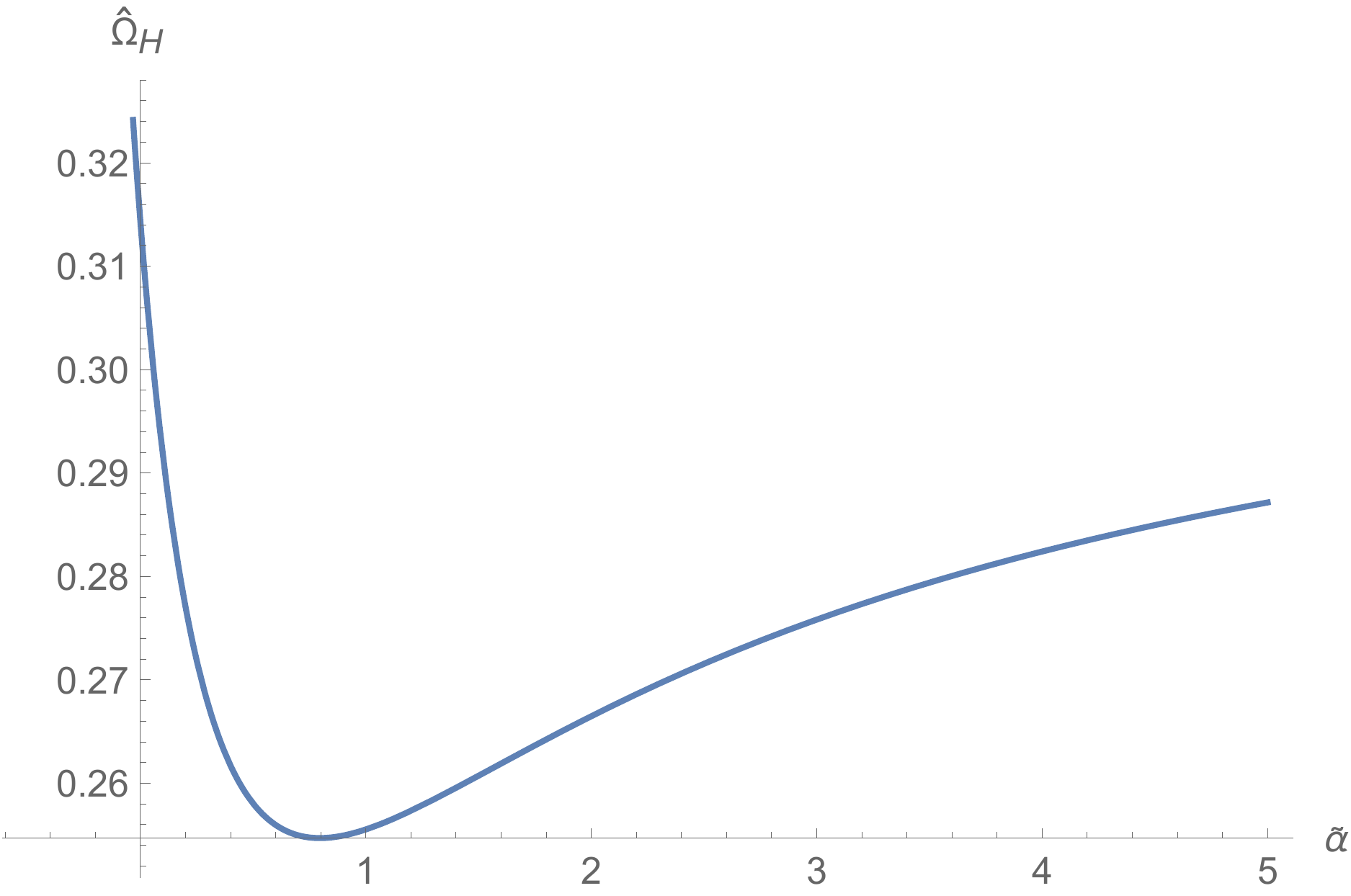}
 \end{center}
 \vspace{-5mm}
 \caption { Dependence of the reduced horizon angular velocity $ \hat{\Omega}_H$ of the EGB black ring on the GB coefficient $\tilde{\alpha}$ with $R=3$. }
 \label{fig:angularvelocity}
\end{figure}

From the expression (\ref{angularvelocity:blackring}) we can easily find that the horizon angular velocity is bounded from below and has a minimum
 at $\tilde{\alpha}=\frac{R^2}{\sqrt{2}(R^2-1)}$, as shown in Fig. \ref{fig:angularvelocity}.
The effect of the GB term on the horizon angular velocity of the black ring is clear: when $-1/2<\tilde{\alpha}<\frac{R^2}{\sqrt{2}(R^2-1)}$, $\hat{\Omega}_H$ decreases with $\tilde{\alpha}$; in contrast, as
$\tilde{\alpha}>\frac{R^2}{\sqrt{2}(R^2-1)}$, $\hat{\Omega}_H$ increases with $\tilde{\alpha}$.
From (\ref{angularvelocity:blackring}) we can see that in the limit $\tilde{\alpha}\to\infty$, we have
\be
\hat{\Omega}_H=\frac{\sqrt{R^2-1}}{R^2},
\ee
identical to the result of the case  $\tilde{\alpha}=0$ \cite{Tanabe1510}.
Under the above condition (\ref{angularvelocity:blackring}) the solution to the equation (\ref{Py:blackring}) is given by
\be\label{Py}
P(y)=P_0+\frac{P_1}{R+y}+\frac{(1+Ry)(1+Ry+2R(R+y)\ln(R+y))}{2R^2(R+y)^2},
\ee
where $P_0$ and $P_1$ are two integration constants. $P_0$ is the redefinition of $r_0$ and $P_1$ comes from the redefinition of the $\phi$ coordinate \cite{Tanabe1510}. The solution (\ref{Py}) is exactly the same as the one in the Einstein
gravity. Actually, the functional forms of $p_z(y)$ and $p_\phi(y)$ take the similar form to the ones in the Einstein gravity, with the dependence on the GB term appearing only in the angular velocity $\hat \Omega_H$ in (\ref{angularvelocity:blackring}).

Summarizing the above results,  we obtain the metric of  the EGB black ring solution at the leading order of the $1/n$ expansion
\bea\label{EGBblackring:leading}
ds^2&=&-\Bigg(1+\frac{1-\Sigma(y)}{8\tilde{\alpha}\hat{\kappa}^2} \Bigg)dv^2+\frac{dvdr}{\hat{\kappa}}\nonumber\\
&&+\frac{(R+y)\sqrt{1-y^2}\,p_v'(y)}{\sqrt{R^2-1}\,p_v(y)}\frac{1-\Sigma(y)}{4\tilde{\alpha}\hat{\kappa}^2}\frac{dvdz}{n}
+\frac{\hat{\Omega}_HR^2(R^2-1)}{(R+y)^2}\frac{1-\Sigma(y)}{4\tilde{\alpha}\hat{\kappa}^2}\frac{dvd\phi}{n}\nonumber\\
&&+\frac{R^2(R^2-1)}{(R+y)^2}\Bigg[1+\frac{1}{n}\Bigg(-\frac{2(1+Ry)}{R^2-1}\ln\sR+\hat{\Omega}_H^2\frac{R^2(R^2-1)}{(R+y)^2}\frac{\Sigma(y)-1}{8\tilde{\alpha}\hat{\kappa}^2} \nonumber\\
&&-\frac{\Big(1+\frac{1+Ry}{4\hat{\kappa}^2R^2}\Big)\Bigg(2\ln\frac{1+\Sigma(y)}{2}+\frac{\pi}{2}-2\arctan \Sigma(y)-\frac{\ln\frac{1+\Sigma(y)^2}{2}}{(1+8\tilde{\alpha}\hat{\kappa}^2)}\Bigg)}{1+4\tilde{\alpha}\hat{\kappa}^2}\Bigg)\Bigg]\frac{d\phi^2}{n}\\
&&+2r^2\frac{\hat{\Omega}_H\sqrt{1-y^2}\sqrt{R^2-1}R^2p_v'(y)}{(R+y)p_v}\frac{\Sigma(y)-1}{8\tilde{\alpha}\hat{\kappa}^2}\frac{dzd\phi}{n^2}+r^2 dz^2+\frac{r^2R^2(1-y^2)}{(R+y)^2}d\Omega_n^2,\nonumber
\eea
where $\Sigma(y)$ is defined in (\ref{Sigmavzphi}) and now becomes
\be
\Sigma(y)=\sqrt{1+\frac{16\tilde{\alpha}\hat{\kappa}^2 p_v(y)}{\sR}}.
\ee
Similar to the black ring in the Einstein gravity \cite{Tanabe1510}, the solution we obtained above breaks down at $R=1$, which means that the large $D$ expansion method cannot capture the behavior of very fat black ring. But both the thin black ring ($R\gg1$) and not-thin black ring ($R>1$) can be described well by the solution.

\subsection{Phase diagram}
With the EGB black ring solution in hand, we can draw the phase diagram of the solution. The thermodynamic quantities for the stationary solutions have been obtained in last section.
From (\ref{blackring:identifications}) it is easy to obtain the
thermodynamic quantities of the EGB black ring. From (\ref{stationarysol:mass}) and (\ref{stationarysol:angularmo}), the mass and the
angular momentum are given by
\be
\mathcal{M}=\frac{n\Omega_n}{8 \mathrm{G}_D}\hat{\mathcal{M}},\qquad \mathcal{J}=\frac{\sqrt{n}\Omega_n}{8 \mathrm{G}_D}\hat{\mathcal{J}},
\ee
where
\be\label{blackring:mass}
 \hat{\mathcal{M}}=\int dy \frac{R(R^2-1)e^{P(y)}}{(R+y)^2\sqrt{1-y^2}}\Bigg(\frac{R\sqrt{1-y^2}}{R+y}\Bigg)^n,
 \ee
 and
 \be\label{blackring:angularmomentum}
 \hat{\mathcal{J}}=\hat{\Omega}_H\int dy  \frac{R^3(R^2-1)^{2}e^{P(y)}}{(R+y)^4\sqrt{1-y^2}}\Bigg(\frac{R\sqrt{1-y^2}}{R+y}\Bigg)^n.
 \ee
From (\ref{stationarysol:entropy}), the entropy of the EGB black ring is given by
\be
S=\frac{\Omega_n \pi}{2G_D} \hat{S},
 \ee
 where
 \be\label{blackring:entropy}
 \hat{S}=\int dy \frac{(R^2+2\tilde{\alpha}(R^2-1))}{(R^2+\tilde{\alpha}(R^2-1))} \frac{R^2\sqrt{R^2-1}e^{P(y)}}{(R+y)^2\sqrt{1-y^2}}\Bigg(\frac{R\sqrt{1-y^2}}{R+y}\Bigg)^n.
 \ee
 From (\ref{stationarysol:tem}) and (\ref{blakcring:kappa}), the temperature of the EGB black ring is given by
 \be\label{blackring:tem}
 T_H=\frac{n}{4\pi}\frac{\sqrt{R^2-1}}{R}\frac{(R^2+\tilde{\alpha}(R^2-1)}{(R^2+2\tilde{\alpha}(R^2-1))}.
 \ee
 The horizon angular velocity is given by
  \be\label{blackring:angularvel}
 \Omega_H=\frac{\hat{\Omega}_H}{\sqrt{n}}=\frac{1}{\sqrt{n}}\frac{\sqrt{R^2-1}}{R^2}\sqrt{\frac{R^4+\tilde{\alpha} R^2(R^2-1)+2\tilde{\alpha}^2(R^2-1)^2}{(R^2+2\tilde{\alpha}(R^2-1))(R^2+\tilde{\alpha}(R^2-1))}}.
 \ee

 \begin{figure}[t]
 \begin{center}
  \includegraphics[width=65mm,angle=0]{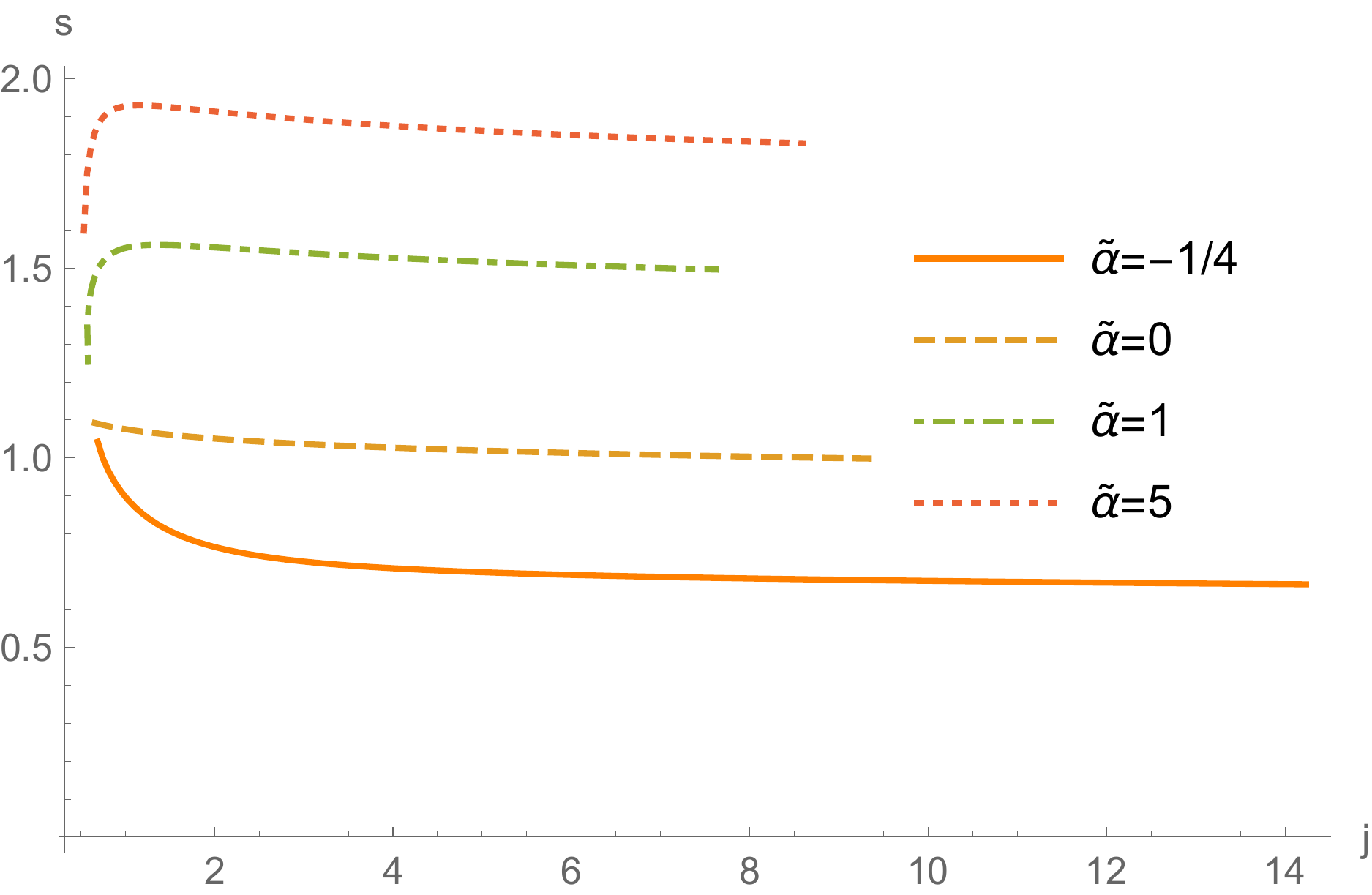}
 \end{center}
 \vspace{-5mm}
 \caption{The $(j, s)$ phase diagrams of the large $D$ EGB black ring solution for $n=30$ ($D=34$). The ring radius $R$ runs from $1.1$ to $20$, so both  the not-thin region and thin region ($R\gg1$) are covered. }
 \label{fig:js}
\end{figure}
 Following \cite{Emparan0708} we define the following dimensionless quantities: the spin $j$,
 the entropy $s$, the angular velocity $\omega_H$ and the temperature $t_H$, by taking the mass as a basic scale
  \be
 j^{n+1}=c_j\frac{\mathcal{J}^{n+1}}{\mathrm{G}_D\mathcal{M}^{n+2}},\qquad
 s^{n+1}=c_s\frac{(4 \mathrm{G}_DS)^{n+1}}{(\mathrm{G}_D\mathcal{M})^{n+2}},
 \ee
 \be
 \omega_H=c_\omega \Omega_H (\mathrm{G}_D\mathcal{M})^{\frac{1}{n+1}},\qquad t_H=c_t(\mathrm{G}_D\mathcal{M})^{\frac{1}{n+1}} T_H,
 \ee
 where the numerical constants  are respectively
 \be
 c_j=\frac{\Omega_{n+1}}{2^{n+5}}\frac{(n+2)^{n+2}}{(n+1)^{(n+1)/2}},\qquad
 c_s=\frac{\Omega_{n+1}}{2(16\pi)^{n+1}}\frac{n^{(n+1)/2}(n+2)^{n+2}}{(n+1)^{(n+1)/2}},
 \ee
 and
 \be
 c_\omega=\sqrt{n+1}\Big(\frac{n+2}{16}\Omega_{n+1}\Big)^{-\frac{1}{n+1}},\qquad
 c_t=4\pi\sqrt{\frac{n+1}{n}}\Big(\frac{n+2}{2}\Omega_{n+1}\Big)^{-\frac{1}{n+1}}.
 \ee
 \begin{figure}[t]
 \begin{center}
  \includegraphics[width=65mm,angle=0]{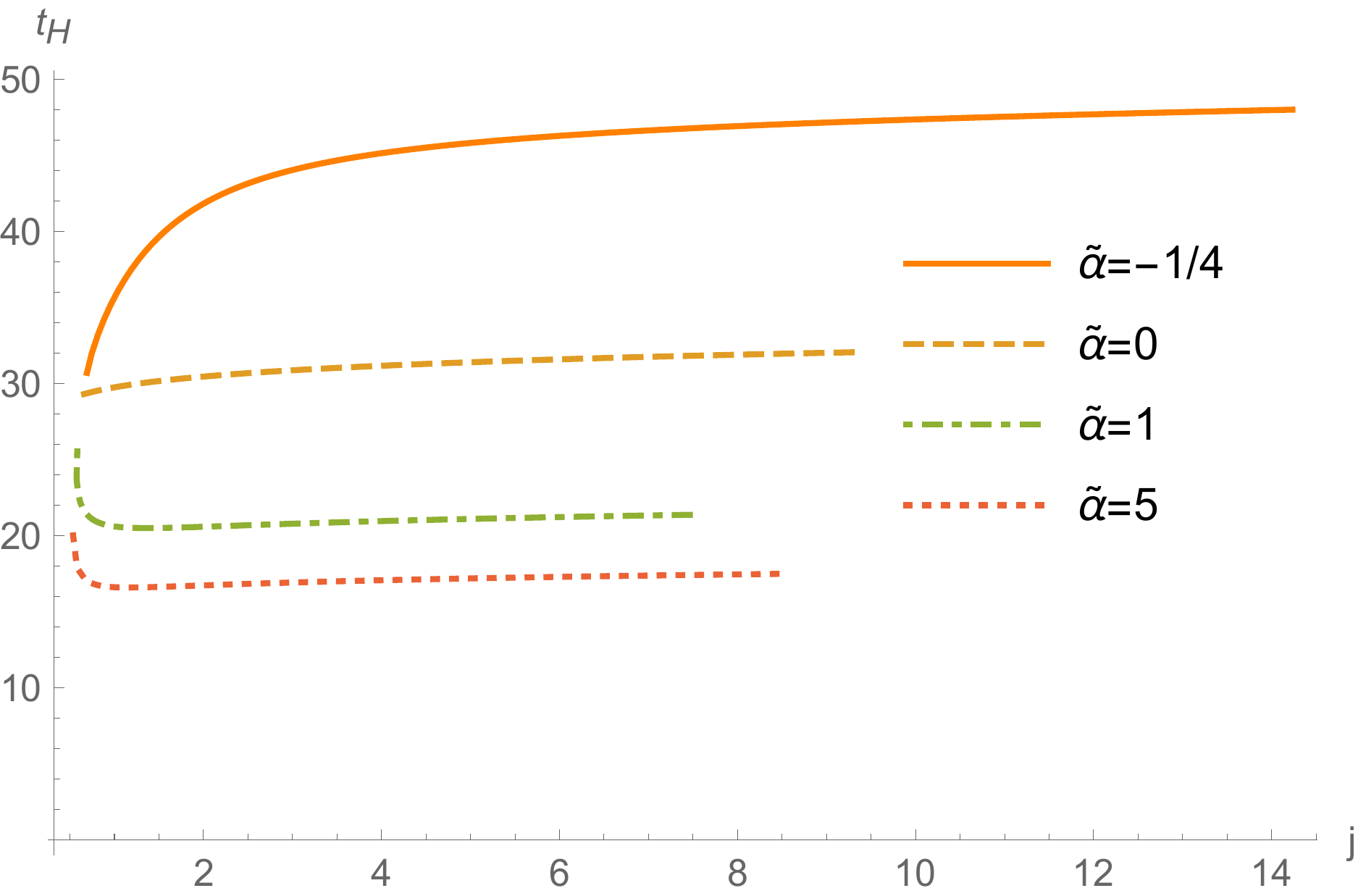}
  \hspace{5mm}
  \includegraphics[width=65mm,angle=0]{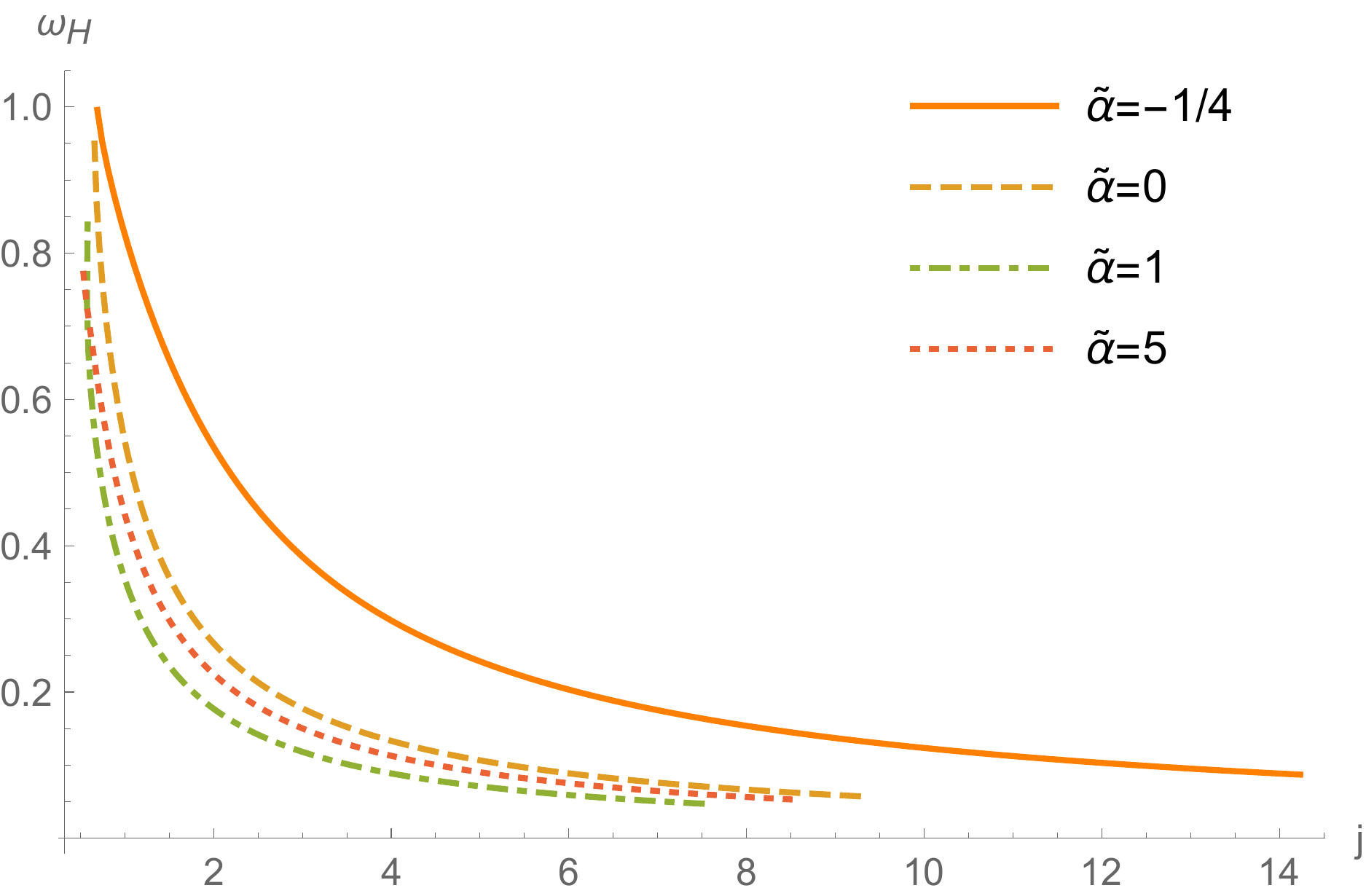}
 \end{center}
 \vspace{-5mm}
 \caption{The $(j, t_H)$ (left panel) and $(j, \omega_H)$ (right panel) phase diagrams of the large $D$ EGB black ring solution. As with Fig. \ref{fig:js}, $n=30$ and $R$ runs from $1.1$ to $20$. }
 \label{fig:jtHjomega}
\end{figure}
 Then we can easily evaluate $j$, $s$, $t_H$ and $\omega_H$ numerically for the large $D$ EGB black ring solution by using (\ref{blackring:mass}), (\ref{blackring:angularmomentum}), (\ref{blackring:entropy}), (\ref{blackring:tem}) and (\ref{blackring:angularvel})
  with $P(y)$ given in (\ref{Py}). For simplicity we set $P_0=P_1=0$.

  In Fig. \ref{fig:js} we show the phase diagrams of $(j, s)$ for the large $D$ EGB black ring solution with different GB coefficient $\tilde{\alpha}$, where $n=30$ and
  $R$ runs from $1.1$ to $20$. We can see easily that in the presence of the GB term, the phase diagrams are  quite different from the ones of the Einstein gravity.
   From (\ref{blackring:entropy}) we know that the entropy of the black ring solution always increases with the GB coefficient. Besides, we find that at small values of $j$ (which correspond to small $R$), $s$ increases with $R$ if $\tilde{\alpha}>0$, in contrast $s$ decreases with $R$ if $-1/2<\tilde{\alpha}<0$.

 We show the phase diagrams of $(j, t_H)$ in the left panel of Fig. \ref{fig:jtHjomega} for the large $D$ EGB black ring solution. From (\ref{blackring:tem})
   we see that  the temperature always decreases with the GB coefficient. The salient difference appearing at small $R$ is due to the fact that $t_H$ decreases with $R$ if $\tilde{\alpha}>0$ but increases with $R$ if $-1/2<\tilde{\alpha}<0$.
 The right panel of Fig. \ref{fig:jtHjomega} is the phase diagrams of $(j, \omega_H)$. On the one hand the horizon angular velocity (\ref{blackring:angularvel}) of the EGB black ring is always lower than that of the Einstein black ring solution if $\tilde{\alpha}>0$ but higher if $-1/2<\tilde{\alpha}<0$. On the other hand, as we mentioned before when $\tilde{\alpha}>\frac{R^2}{\sqrt{2}(R^2-1)}$, the horizon angular velocity increases with $\tilde{\alpha}$.

\section{Quasinormal modes and non-linear evolution of EGB black ring} \label{section4}

\subsection{Quasinormal modes}
We would like to study the quasinormal modes (QNMs) of the EGB black ring solution we obtained before. The QNMs are obtained by the
perturbation analysis of the effective equations around the stationary  black ring solution. We consider the following perturbation ansatz
\bea\label{perturbation:EGBblackringpv}
p_v(v,y,\phi)&=&e^{P(y)}\Big(1+\epsilon e^{-i\omega v} e^{im\phi} F_v(y)\Big),\\\label{perturbation:EGBblackringpz}
p_z(v,y,\phi)&=&-\frac{(R+y)\sqrt{1-y^2}}{\sqrt{R^2-1}}e^{P(y)}P'(y)\Big(1+\epsilon e^{-i\omega v} e^{im\phi} F_z(y)\Big),\\\label{perturbation:EGBblackringpphi}
p_\phi(v,y,\phi)&=& \sqrt{\frac{R^4+\tilde{\alpha} R^2(R^2-1)+2\tilde{\alpha}^2(R^2-1)^2}{(R^2+2\tilde{\alpha}(R^2-1))(R^2+\tilde{\alpha}(R^2-1))}}\frac{(R^2-1)^{3/2}}{(R+y)^2}e^{P(y)}\Big(1+\epsilon e^{-i\omega v} e^{im\phi} F_\phi(y)\Big).\nonumber\\
\eea
As in \cite{Tanabe1510}, in this paper we consider the case $m=\mc O(1)$, as we can see in the following the only instability in this case is the GL-like instability.
There is another choice for $m$, i.e. $m=\mc O(1/\sqrt{n})$, from (\ref{phiandPhi}) this amounts to $m_\Phi=\mc O(1)$.
 A new elastic-type instability occurs in this case, as first found by the numerical study in \cite{Figueras1512}, and then by the large $D$ study in \cite{Tanabe16050811}. To investigate the elastic instability, the $\mc O(1/n^2)$ corrections to the
effective equations are needed, which is more complicated for the EGB theory, so in this paper we only consider the GL-like instability.

Since we consider the case $m=\mc O(1)$, the above ansatz  for the perturbations around the stationary solutions is appropriate. Substituting the above perturbations into the effective equations, and taking into account that at the pole
 $y=-1/R$, $F_v(y)$ behaves like
 \be\label{blackringBC}
 F_v(y)\propto(1+Ry)^\ell,
 \ee
 such that the perturbation fields may have regular solutions.

Due to the special form of the metric ansatz, the vector-type perturbation can be studied only for the slowly rotating EGB black hole, as we will see in the next section. For the EGB black ring and the slowly boosted EGB black string, the momentum density other than $p_z$ is needed to study the vector-type perturbation. Here we only consider the most interesting scalar-type gravitational perturbation. The QNM condition for the scalar-type gravitational perturbation is given by
\bea\label{QNMcondition:EGBblackring}
&&\omega^3+\frac{i\,\omega^2\sqrt{R^2-1}}{2R^3}\Bigg(3 m^2 (
\zeta^2+1)+6 i m R \zeta+R^2 (\zeta^2+1) (3 \ell -2)\Bigg)\nonumber\\
&&-\frac{\omega\,(R^2-1) }{2R^6}
\Bigg(m^4(\zeta^4+4\zeta^2+1)+6 i m^3R \zeta (\zeta^2+1)+2m^2 R^2(\zeta^4+4 \zeta^2+1) \ell\nonumber\\
&&+ 2 i mR^3 \zeta (\zeta^2+1) (3 \ell -2)+R^4(\zeta^4+4 \zeta^2+1)(\ell -1) \ell-2m^2 R^2(\zeta^4+6\zeta^2+1)\Bigg)\nonumber\\
&&-\frac{i\zeta(R^2-1)^{3/2} }{2R^9}\Bigg[ m^6\zeta(1 + \zeta^2)+i m^5R (\zeta^4+4 \zeta^2+1)+3 m^4R^2 \zeta (\zeta^2+1) (\ell -2)\nonumber\\
&&+2 i m^3R^3 (\zeta^4+4 \zeta^2+1) (\ell -1)+R^4m^2 \zeta(\zeta^2+1)(3 \ell ^2-7 \ell +4)\nonumber\\
&&+im R^5(\zeta^4+4 \zeta^2+1) (\ell -1) \ell+R^6 \zeta (\zeta^2+1) (\ell -1) \ell ^2\Bigg]=0,
\eea
where we have used $\zeta$ to denote
\be
\zeta=\sqrt{\frac{2\tilde{\alpha}^2-R^2\tilde{\alpha}(1+4\tilde{\alpha})+R^4(1+\tilde{\alpha}+2\tilde{\alpha}^2)}{2\tilde{\alpha}^2-R^2\tilde{\alpha}(3+4\tilde{\alpha})+R^4(1+3\tilde{\alpha}+2\tilde{\alpha}^2)}}.
\ee
\begin{figure}[t]
 \begin{center}
  \includegraphics[width=65mm,angle=0]{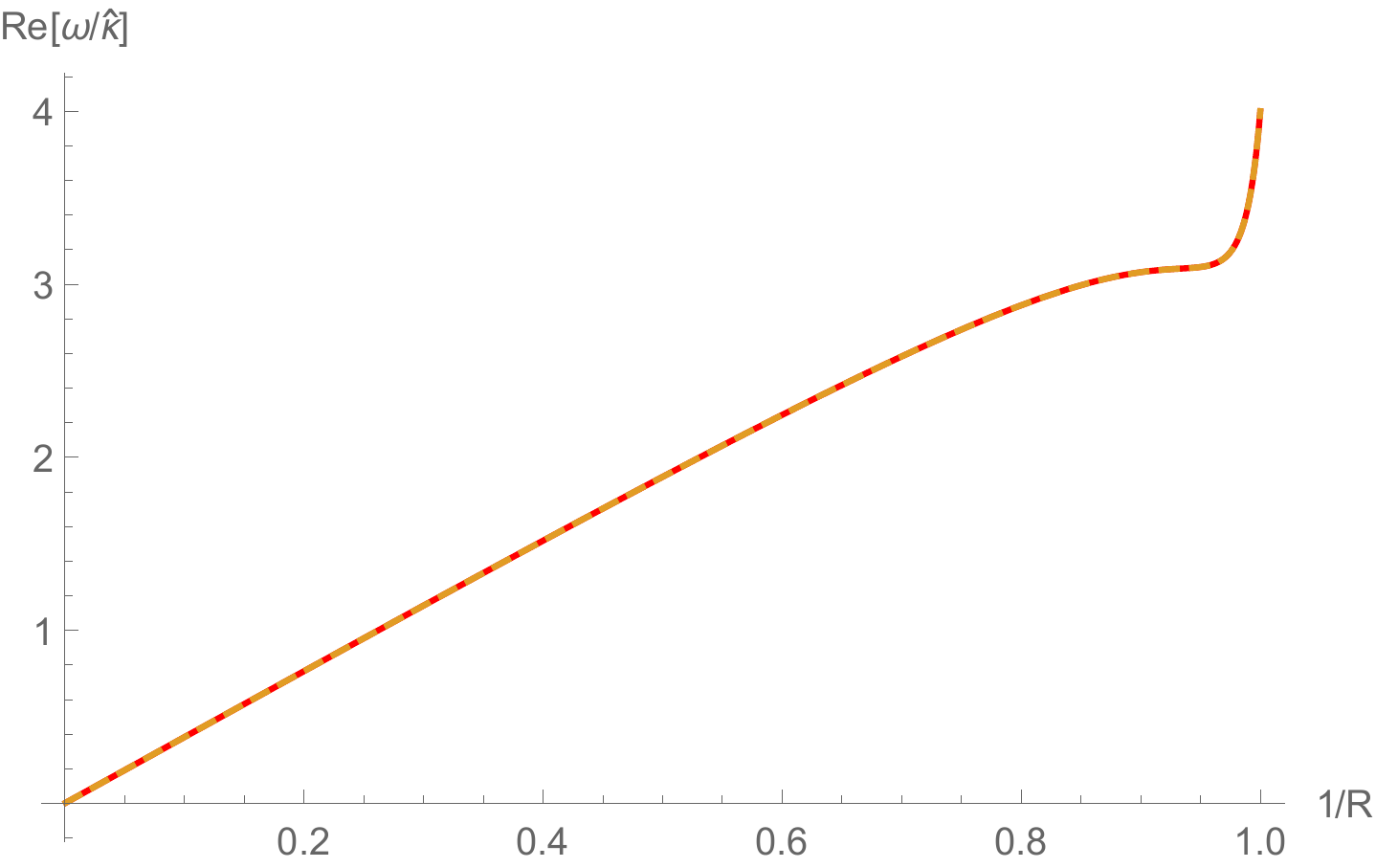}
  \hspace{5mm}
  \includegraphics[width=65mm,angle=0]{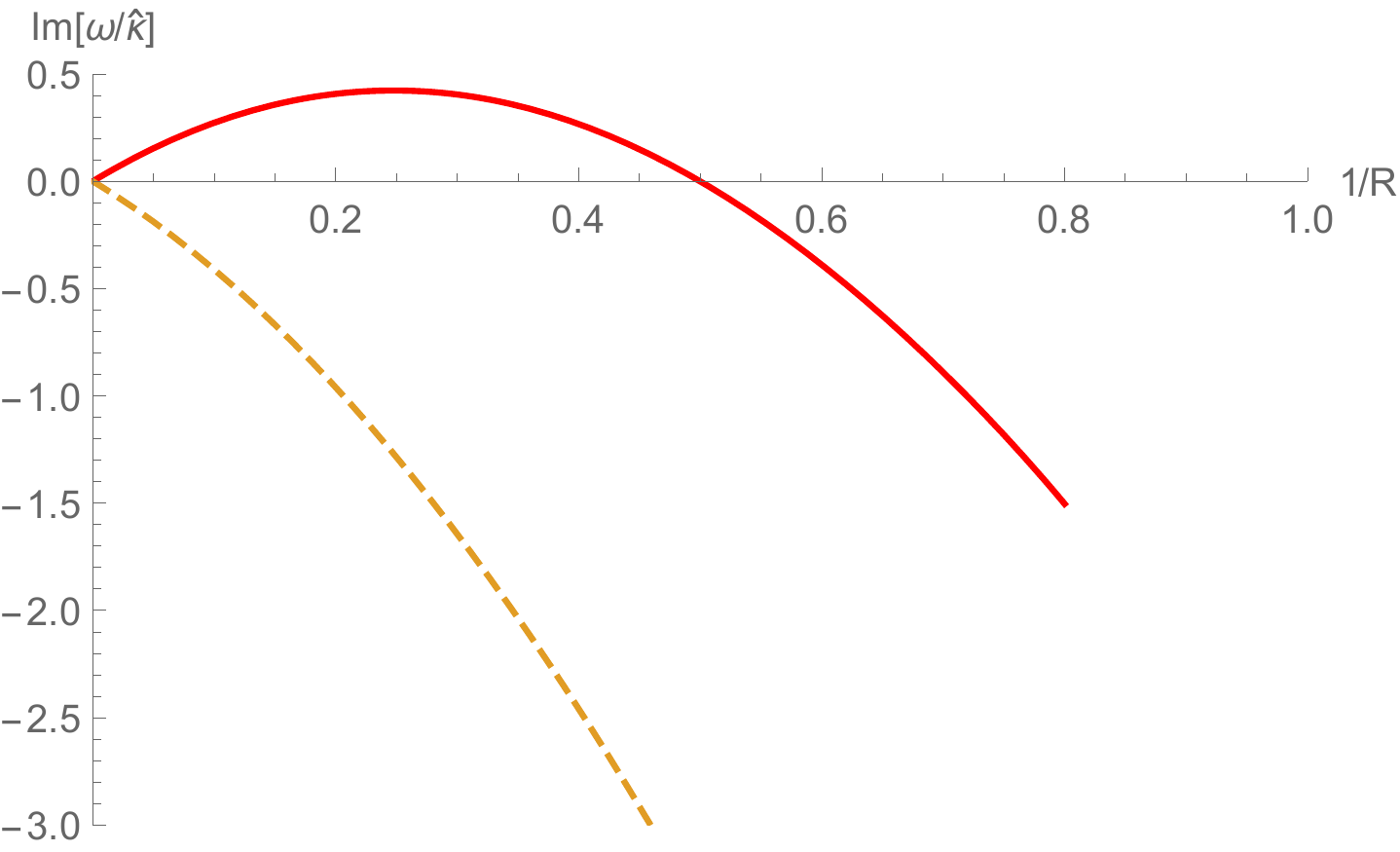}
 \end{center}
 \vspace{-5mm}
 \caption { The  frequencies $\omega_{+}^{(\ell=0)}$ (solid line) and $\omega_{-}^{(\ell=0)}$ (dashed line) of the scalar-type quasinormal modes of the EGB black ring with $\ell=0$, $m=2$ and $\tilde{\alpha}=2$ . The real part and the imaginary part are depicted in the left and right panel respectively, both of which are in unit of the reduced surface gravity $\hat{\kappa}$. $\mathrm{Im}[\omega_{+}^{(\ell=0)}]$ is positive for $R>2$ which signals the GL-like instability.}
 \label{fig:QNMofblackring}
\end{figure}
In the limit $\tilde{\alpha}\to0$, i.e. $\zeta\to1$, the above equation is reduced to the QNM condition for the black ring in
the Einstein gravity \cite{Tanabe1510}. Qualitatively, the QNMs are quite similar to the  ones of the black ring in the Einstein gravity. For $\ell\neq0$ and $m\neq0$, all modes are  stable.
For $\ell=0$ and $m\neq0$, the QNM condition can be solved by
 \be
 \omega_0^{(\ell=0)}=\frac{\sqrt{R^2-1}}{2R}\Bigl[2\hat{m}\zeta+2i(1+\zeta^2)-i\hat{m}^2(1+\zeta^2)\Bigl],
 \ee
 and
 \be\label{QNM:EGBblackring}
 \omega_{\pm}^{(\ell=0)}=\frac{\hat{m}\sqrt{R^2-1}}{2R}
 \Bigl[2\zeta-i\hat{m}(1+\zeta^2)\pm i\sqrt{4\zeta^2+\hat{m}^2(1-\zeta^2)^2} \Bigl],
 \ee
 where $\hat{m}=m/R$. Similar to the case of the large $D$ black ring in the Einstein gravity, we have three QNMs.
 The mode with frequency $\omega_0^{(\ell=0)}$ should be a gauge mode, as in the large radius limit
 $R\gg1$, the counterpart of the boosted EGB black string is a gauge mode. In contrast, as we will see below,
 the other two modes $\omega_{\pm}^{(\ell=0)}$ can be understood as the QNMs of the boosted EGB black string in the same limit.
 A typical result of the real and imaginary part of the frequencies $\omega_{\pm}^{(\ell=0)}$ is
 plotted in Fig. \ref{fig:QNMofblackring}, with $m=2$ and $\tilde{\alpha}=2$. It is easy to see that when
 $R>m$, $\omega_{+}^{(\ell=0)}$ has a positive imaginary part, indicating the GL-like  instability.
 It is interesting to note that the critical radius at which $\mathrm{Im}[\omega_{+}^{(\ell=0)}]$ vanishes
 is independent of the GB coefficient. This is reminiscent of the property of the EGB black string that
 the wavenumber of the threshold mode of the GL instability is also independent of the GB coefficient \cite{Chen1707}.

\begin{figure}[t]
 \begin{center}
  \includegraphics[width=65mm,angle=0]{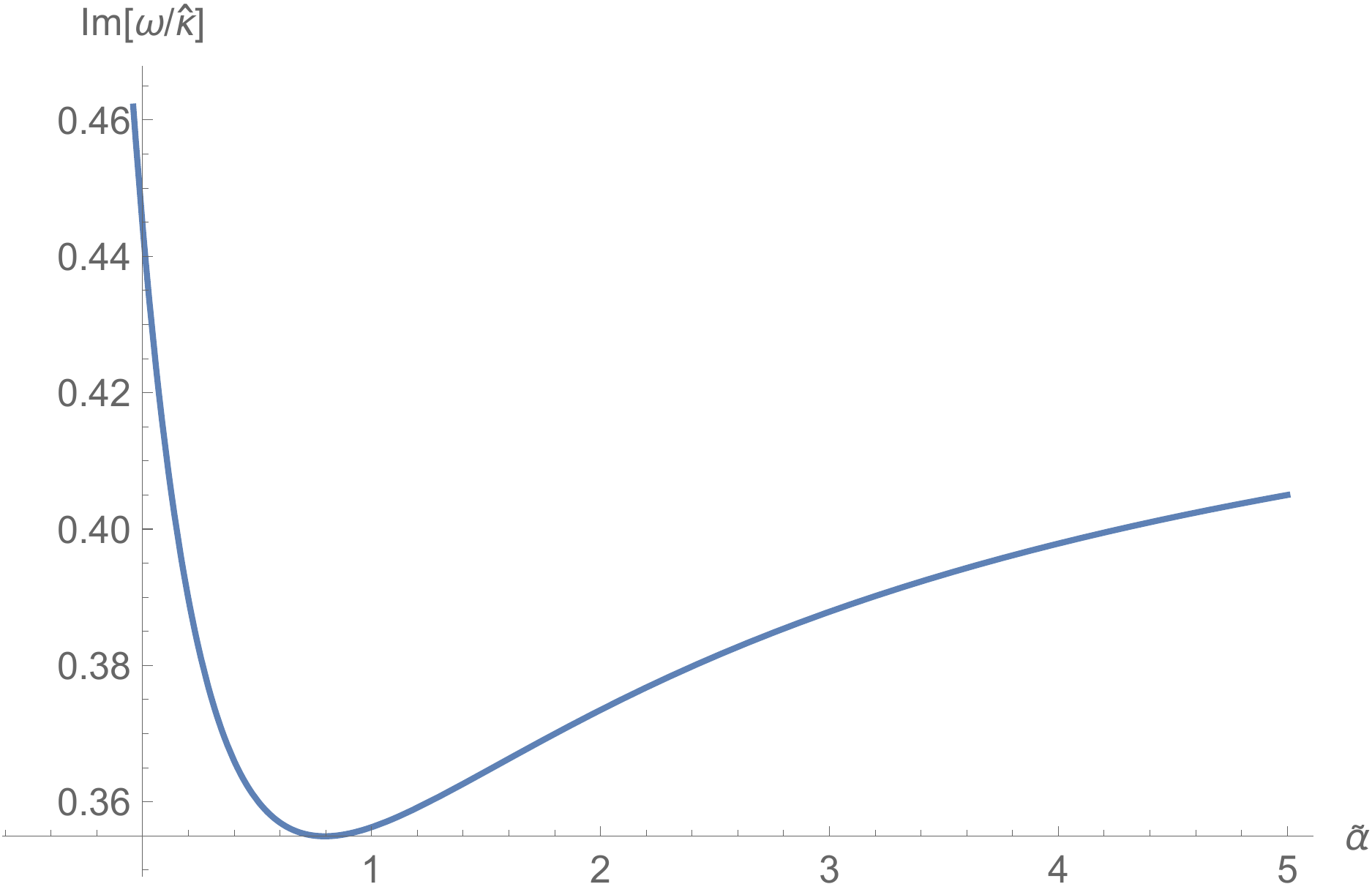}
 \end{center}
 \vspace{-5mm}
 \caption { Dependence of  Im$ [{\omega_+^{\ell=0}/\hat{\kappa}}]$ of the EGB black ring on the GB coefficient $\tilde{\alpha}$ with $R=3$. }
 \label{fig:Imwrtalpha}
\end{figure}

From (\ref{QNM:EGBblackring}) we find that the dependence of the instability on the GB coefficient is the
same as that of the horizon angular velocity of the black ring, as shown in Fig. \ref{fig:Imwrtalpha}. When $-1/2<\tilde{\alpha}<\frac{R^2}{\sqrt{2}(R^2-1)}$, the increase of $\tilde{\alpha}$ alleviates the instability. In contrast, as
$\tilde{\alpha}>\frac{R^2}{\sqrt{2}(R^2-1)}$ the increase of $\tilde{\alpha}$ enhances the instability. A
similar phenomena has been found for the EGB black string \cite{Chen1707}, as we will show below they are equivalent in the large
radius limit.

It is known that for the black rings in the Einstein gravity \cite{Tanabe1510} and the Einstein-Maxwell theory \cite{Chen1702}, the real part of  $\omega_{+}^{(\ell=0)}$ always saturates the superradiant condition, i.e.
\be
 \text{Re}\Big[ \omega_{+}^{(\ell=0)}\Big]=m\hat{\Omega}_H.
 \ee
 For the black ring in EGB theory the same thing occurs with $\hat{\Omega}_H$ given by (\ref{angularvelocity:blackring}).
 Thus at the leading order of the large $D$ expansion, the coincidence of the onsets of  the dynamical instability and the superradiant
 condition seems to be a robust feature for both the Einstein gravity and the EGB theory. By considering the $1/n$ correction, the above
 relation is broken \cite{Tanabe1510}, and we expect this happens for the EGB black ring as well.

For the axisymmetric modes $m = 0$, the QNM condition is solved by
\be
\omega_0^{(m=0)}=-i\frac{\sqrt{R^2-1}}{2R}\ell(1+\zeta^2),
\ee
and
\be\label{axisymQNM:EGBblackring}
\omega_{\pm}^{(m=0)}=\frac{\sqrt{R^2-1}}{2R}\Bigg[-i \, (\ell-1)(1+\zeta^2)\pm\sqrt{(1-\ell)
\Big(\ell(\zeta^2-1)^2-(\zeta^2+1)^2\Big)}\Bigg].
\ee
 As discussed in \cite{ Chen1702, Emparan1402}, the mode $\omega_0^{(m=0)}$ should correspond to the vector-type gravitational perturbations of the EGB black hole in the large
 radius limit.
It was found that for the black rings in the Einstein gravity \cite{Tanabe1510} and the Einstein-Maxwell theory \cite{Chen1702}, the frequencies for the axisymmetric QNMs are  the same as the ones of the large $D$ Schwarzschild and Reissner-Nordstrom  black holes up to a  multiplicative factor characterizing the finiteness of the ring. However, it is easy to see that for the EGB black ring this does not hold any more. This is because
unlike the case in the Einstein gravity, although for the $m=0$ modes the perturbations do not have interaction along $\phi$ direction,
from  (\ref{perturbation:EGBblackringpv}), (\ref{perturbation:EGBblackringpz}) and (\ref{perturbation:EGBblackringpphi}) we can see that the situation is still different from the one of
the  $D-1$ dimensional EGB black hole. However, as we will see in the next section, for the slowly boosted EGB black string  the QNMs of the axisymmetric
perturbations are the same as the ones of the EGB black hole \cite{Chen1703}.

 Like the case without the GB term, the axisymmetric perturbations do not show any instability at the leading order of the $1/n$ expansion.
 However, it has been found that for a fat black ring in five dimensions the axisymmetric perturbations are unstable \cite{Elvang0608076, Figueras1107,Santos1503, Figueras1512}. The reason why the large $D$ expansion can not capture this feature is not clear at present.

\subsection{Black String limit }

It is well known that for the Einstein gravity when the ring radius $R\gg1$, that is the  black ring is very thin, the black ring is identical to a critically boosted black string \cite{Emparan0708}, in which   the boost velocity is  fixed. In the following we show this occurs for the EGB theory as well. In the large ring radius limit, from (\ref{Py}) $P(y)$ behaves like
\be
P(y)=\mc O(1/R),
\ee
by setting $P_0=P_1=0$. The black ring solution (\ref{EGBblackring:leading}) in the large radius limit becomes
\bea\label{largeRlimitmetric}
ds^2&=&-\Bigg(1+\frac{1-\Sigma(y)}{2\tilde{\alpha}} \Bigg)dv^2+2dvdr+\sqrt{\frac{1+\tilde{\alpha}+2\tilde{\alpha}^2}{1+3\tilde{\alpha}+2\tilde{\alpha}^2}}\frac{1-\Sigma(y)}{\tilde{\alpha}}\frac{dvdx}{n}\nonumber\\
&&+\Bigg[1+\frac{1}{n}\Bigg(\frac{1+\tilde{\alpha}+2\tilde{\alpha}^2}{1+3\tilde{\alpha}+2\tilde{\alpha}^2}\frac{\Sigma(y)-1}{2\tilde{\alpha}} -\frac{2\ln\frac{1+\Sigma(y)}{2}+\frac{\pi}{2}-2\arctan \Sigma(y)-\frac{\ln\frac{1+\Sigma(y)^2}{2}}{(1+2\tilde{\alpha})}}{1+\tilde{\alpha}}\Bigg]\frac{dx^2}{n}\nonumber\\
&&+r^2 dz^2+r^2\sin^2z d\Omega_n^2,\nonumber\\
\eea
where the string direction $dx$ is identified by $dx=Rd\Phi=Rd\phi/\sqrt{n}$.  Under such large radius limit, the above expression is equivalent to the metric of the
large $D$ boosted EGB black string \cite{Chen1707} with the boost velocity $\beta$ being
\be\label{largeRlimitboost}
\sinh\beta=\frac{1}{\sqrt{n}}\sqrt{\frac{1+\tilde{\alpha}+2\tilde{\alpha}^2}{1+3\tilde{\alpha}+2\tilde{\alpha}^2}}.
\ee
It is easy to see that this boost velocity equals the horizon angular velocity $\Omega_H$ (\ref{angularvelocity:blackring}) (note that
 $\Omega_H=\hat{\Omega}_H/\sqrt{n}$) in the large radius limit. Using the boost relation we can associate the large radius limit of the
 QNMs of the  EGB black ring with the ones of the EGB black string. The QNMs of the  EGB black string were
 obtained in \cite{Chen1707} by using the large $D$ effective theory. For the S-wave sector ($\ell=0$) of the gravitational perturbation, the
 QNMs are given by
\be
\omega_{\pm}^{st}=-i\hat{k}^2\frac{1+2\tilde{\alpha}+2\tilde{\alpha}^2}{(1+\tilde{\alpha})(1+2\tilde{\alpha})}\pm\frac{i \hat{k}\sqrt{1+4\tilde{\alpha}+(7+\hat{k}^2)\tilde{\alpha}^2+8\tilde{\alpha}^3+4\tilde{\alpha}^4}}{(1+\tilde{\alpha})(1+2\tilde{\alpha})},
\ee
where $\hat{k}=k/\sqrt{n}$ is the quantum number with respect to $\phi$. Performing the following boost transformation
\be
dv\to \mathrm{cosh}\,\alpha\, dv-  \mathrm{sinh}\,\alpha\,dx,\qquad dx\to \mathrm{cosh}\,\alpha\, dx-  \mathrm{sinh}\,\alpha\,dv,
\ee
with the  boost velocity given by (\ref{largeRlimitboost}), we obtain the  QNMs of the boosted EGB black string
\be
\omega_{\pm}^{bst}=k\sinh\,\alpha+\omega_{\pm}^{st} \cosh\,\alpha.
\ee
The above result is identical to the QNMs of the EGB black ring (\ref{QNM:EGBblackring}) in the large radius limit if we
identify $\hat{k}=\hat{m}$. In the next section we will give a direct derivation of the QNMs of the
slowly boosted EGB black string from the effective equations and show that they are related to the ones of \cite{Chen1707}
 by a simple boost transformation.

\subsection{Non-linear evolution}
As we discussed before, the thin EGB black ring at large $D$ suffers from the GL-like instability for the non-axisymmetric perturbations. It would be interesting to
 explore the fate of the instability in the non-linear regime at late time. For the Einstein black string it was found \cite{Sorkin0402216} that above a critical dimension $D^*\simeq 13.5$ the weakly non-uniform black strings have larger horizon areas than the uniform ones, which implies that the non-uniform black string could be the possible end point of the non-linear evolution.  Using the large $D$ effective theory
 \cite{Emparan1506}, the non-linear evolution of the black string instability was demonstrated and it was shown that
at late time the unstable black strings in a large enough number of dimensions end at stable non-uniform black strings, which supports the conjecture in \cite{Sorkin0402216}. Furthermore, like the case in the Einstein gravity,  the thin EGB black strings at large $D$ are unstable to
developing inhomogeneities along their length, and at late time they asymptote to the stable non-uniform black strings \cite{Chen1707}.

As we mentioned in the introduction, numerical study in five dimensions \cite{Figueras1512} gives strong evidence that the non-linear evolution of the GL-like instability does not stop at any stable configuration but leads to the pinch-off of the ring. For the black ring in higher dimensions, due to the similarity between the black ring and the black string, it was conjectured that the
end point of the black ring instability should be a large $D$ non-uniform black ring   \cite{Tanabe1510}.\footnote{Due to the existence of the gravitational
wave emissions, the non-uniform black ring solution is not stationary in a much longer timescale, thus the rigidity theorem \cite{Hollands0605106} is not broken.} In this subsection by numerically solving the effective equations for the  EGB black ring near $y=-1/R$, we give a relative strong evidence to support the conjecture in \cite{Tanabe1510}.

\subsubsection{Einstein black ring}

Firstly, we study the non-linear evolution of the black ring in the Einstein gravity by numerically solving the effective equations.
The effective equations for the black ring in the Einstein gravity can be derived by combining the general effective equations (\ref{Effeq1}), (\ref{Effeq2}) and (\ref{Effeq3}) with the identifications (\ref{blackring:identifications}) and taking $\tilde{\alpha}\to0$ . They are given by
\be\label{Effeq1:BR}
\partial_vp_v+\frac{(R+y) (R y+1)}{R \sqrt{R^2-1}} \partial_yp_v
-\frac{(R+y)^2 }{R \left(R^2-1\right)^{3/2}}\partial_\phi^2p_v
+\frac{(R+y)^2}{R^2 \left(R^2-1\right)} \partial_\phi p_\phi
+\frac{(R y+1)}{R \sqrt{1-y^2}} p_z=0,
\ee
\bea\label{Effeq2:BR}
&&\partial_vp_z+\frac{(R+y) (R y+1)}{R \sqrt{R^2-1}}\partial_y p_z
-\frac{\sqrt{1-y^2} (R+y)}{R}\partial_y p_v-\frac{(R+y)^2}{R \left(R^2-1\right)^{3/2}}\partial_\phi^2 p_z
+\frac{(R y+1)}{R \sqrt{1-y^2}}\frac{p_z^2}{ p_v}\nonumber\\
&&\quad +\frac{\sqrt{1-y^2} (R+y) p_v}{R^2-1}+\frac{\left(R^2 \left(1-2 y^2\right)-2 R y+y^2-2\right)p_z}{R \sqrt{R^2-1} \left(y^2-1\right)}
-\frac{\sqrt{1-y^2} (R+y)^2 p_\phi^2}{R^3 \left(R^2-1\right) p_v}\\
&&\quad +\frac{(R+y)^2}{R^2 \left(R^2-1\right)}\partial_\phi\Big[\frac{p_z p_\phi}{p_v}\Big]
+\frac{2 \sqrt{1-y^2} (R+y)^2 }{R^2 \left(R^2-1\right)^{3/2}}\partial_\phi p_\phi=0\nonumber,
\eea
\bea\label{Effeq3:BR}
&&\partial_v p_\phi+\frac{(R+y) (R y+1)}{R \sqrt{R^2-1}}\partial_y p_\phi
-\frac{(R+y)^2 }{R \left(R^2-1\right)^{3/2}}\partial_\phi^2 p_\phi
+\frac{ (R+y)^2 }{R^2 \left(R^2-1\right)}\partial_\phi\Big[\frac{ p_\phi^2}{p_v}\Big]\nonumber\\
&&\quad -\frac{\left(R^2+2 R y+1\right)}{R^2-1}\partial_\phi p_v
+\frac{(R y+1)}{R \sqrt{1-y^2}}\frac{p_z p_\phi}{p_v}+\frac{2 (R y+1)}{R \sqrt{R^2-1}}p_\phi=0.
\eea
By solving these equations numerically, we can study the non-linear evolution of the black ring.
To achieve this end we need to specify the boundary conditions in $\phi$ and $y$ directions. It is
obvious that the solution should be periodic in $\phi$ direction.
However, the boundary condition in $y$ direction is not clear. Since we expect to obtain the non-uniform black
ring solution from these equations, the solution is expected to be inhomogeneous along the $\phi$ direction no matter what
$y$ takes.  Thus we might be able to solve the effective equations without involving the boundary condition in $y$ direction if $y$ takes
a specific value. Indeed, near $y=-1/R$ the equations (\ref{Effeq1:BR}) and (\ref{Effeq3:BR}) become
\be\label{Eq1:specialcase}
\partial_vp_v
-\frac{\sqrt{R^2-1}}{R^3}\partial_\phi^2p_v
+\frac{R^2-1}{R^4} \partial_\phi p_\phi=0,
\ee
\be\partial_v p_\phi
-\frac{\sqrt{R^2-1}}{R^3}\partial_\phi^2 p_\phi
+\frac{ (R^2-1) }{R^4}\partial_\phi\Bigg[\frac{ p_\phi^2}{p_v}\Bigg]
-\partial_\phi p_v=0.\label{Eq2:specialcase}
\ee
Clearly, $p_z$ is decoupled from these two equations. In this case although (\ref{Effeq2:BR}) is still complicated,  we can solve $p_v$ and $p_\phi$ via the above two equations without the information
of $p_z$.   In this way, the inhomogeneity along $\phi$ direction is clearly embodied. In the large radius limit, taking into account that $\partial_\phi= R \partial_x $ and
 $p_\phi=R p_x$ ($x$ denotes the string direction), the above equations are exactly the effective equations for the black string in the Einstein
gravity \cite{Emparan1506}. But as we explained in the previous subsection, in this case these equations actually describe slowly boosted black string, thus
the initial condition used to simulate these equations should be different. In particular,  even for the uniform black string $p_x$ should not be zero due to the existence of the angular momentum.
For a finite $R$  solving the above equations numerically should lead to non-trivial result.

From the perturbation analysis we know that the instability occurs when $\hat{m}=m/R<1$, so if we fix the value of $m$ then the following
 behavior of the initial perturbation is completely determined by the value of $R$.  For $\phi$ direction, it is compactified, $\phi\in[-\pi/m, \pi/m]$, which
 in the large radius limit is equivalent to $x\in[-L/2, L/2]$, where $L$ denotes the periodicity of the string direction.
\begin{figure}
 \begin{center}
  \includegraphics[width=34mm,angle=0]{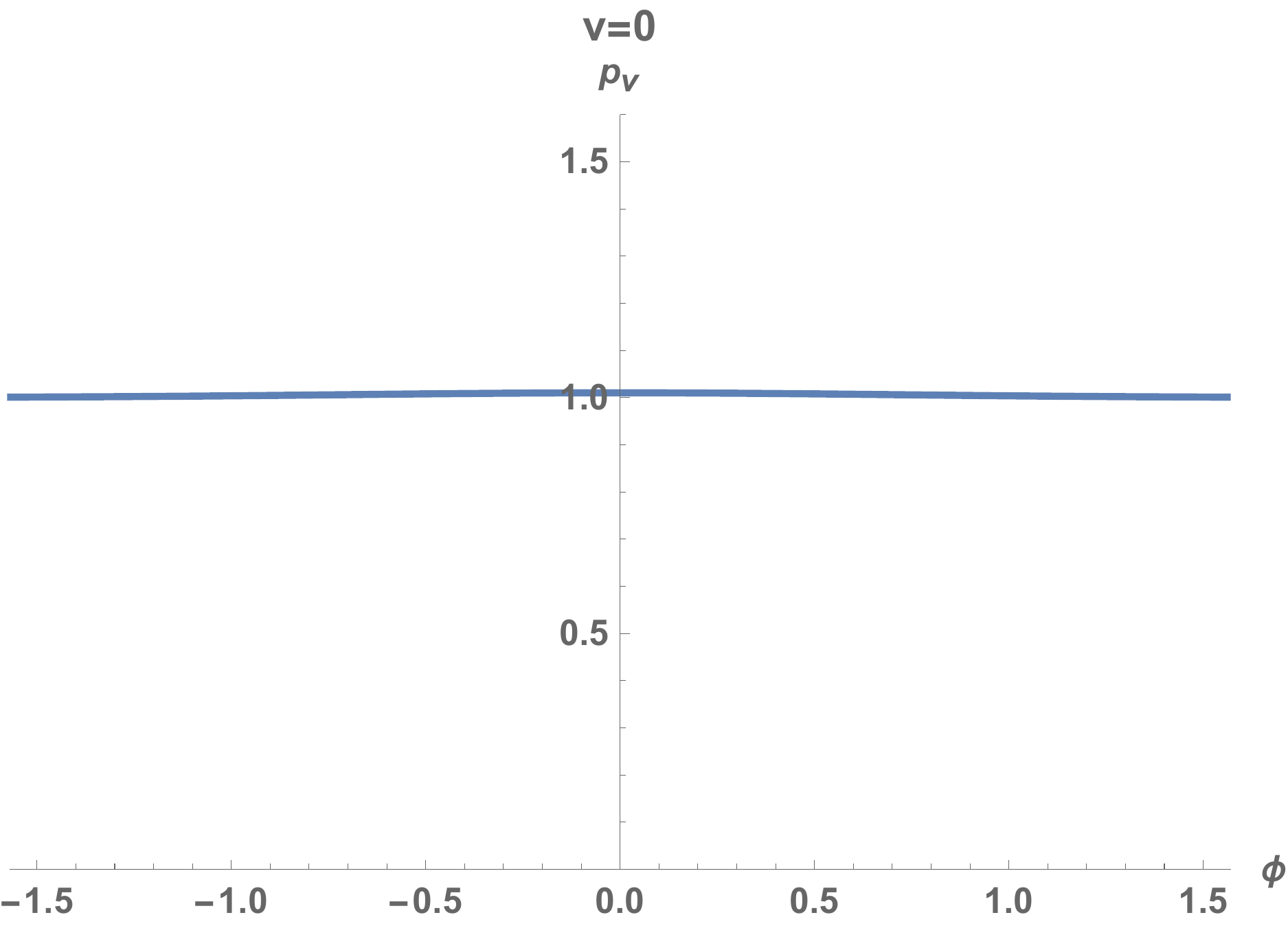}
 \hspace{1mm}
  \includegraphics[width=34mm,angle=0]{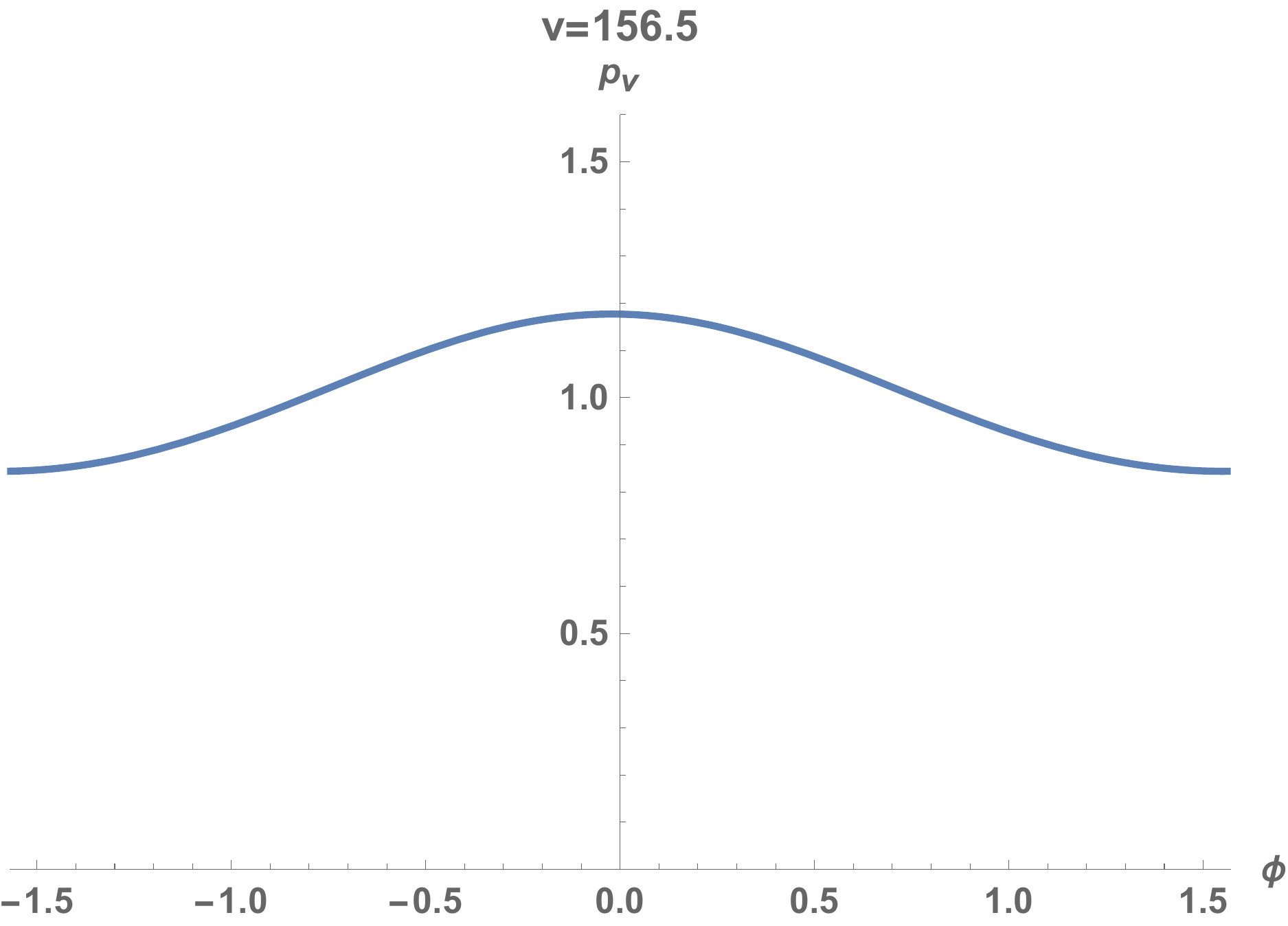}
   \hspace{1mm}
  \includegraphics[width=34mm,angle=0]{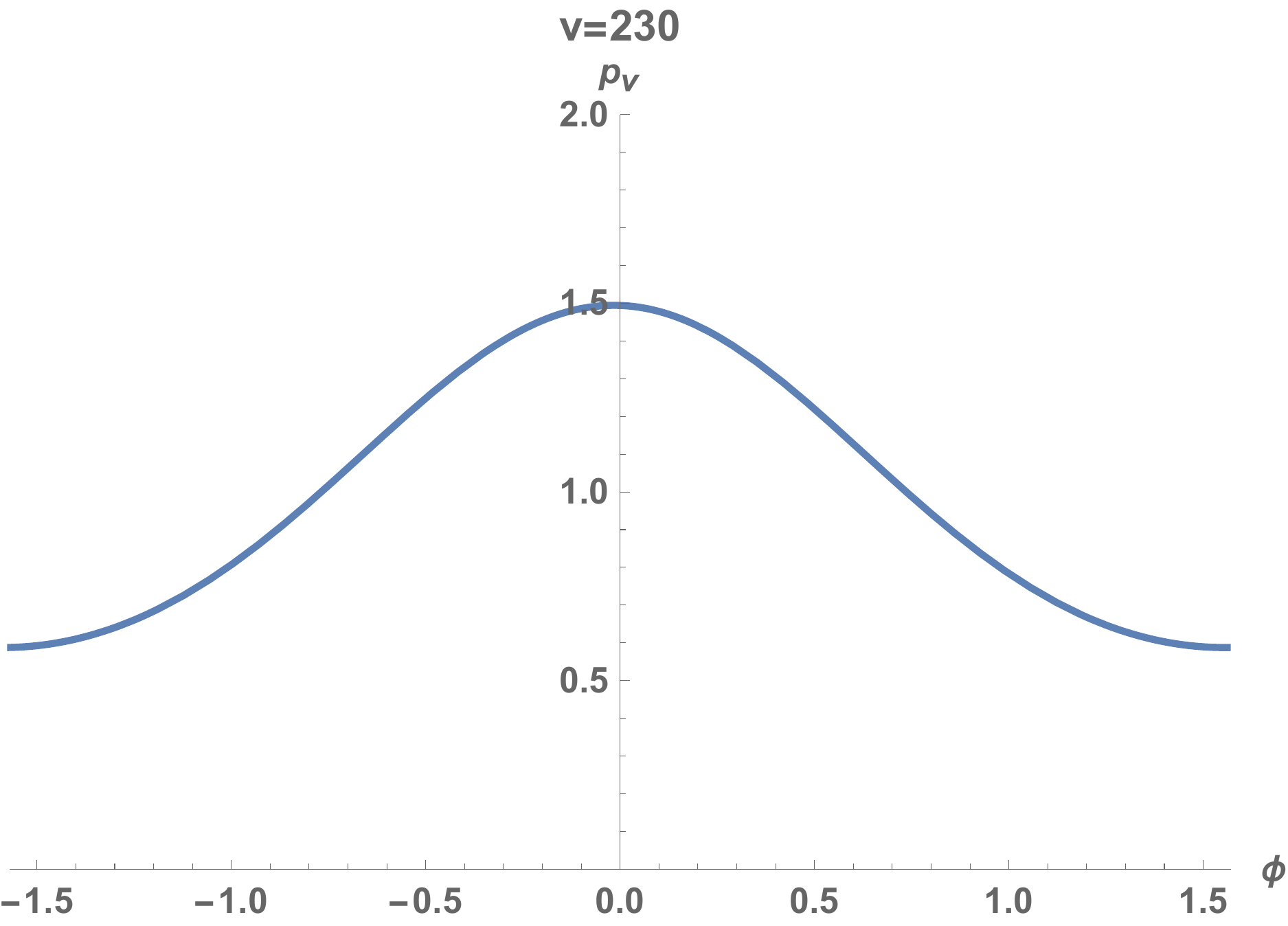}
  \hspace{1mm}
  \includegraphics[width=34mm,angle=0]{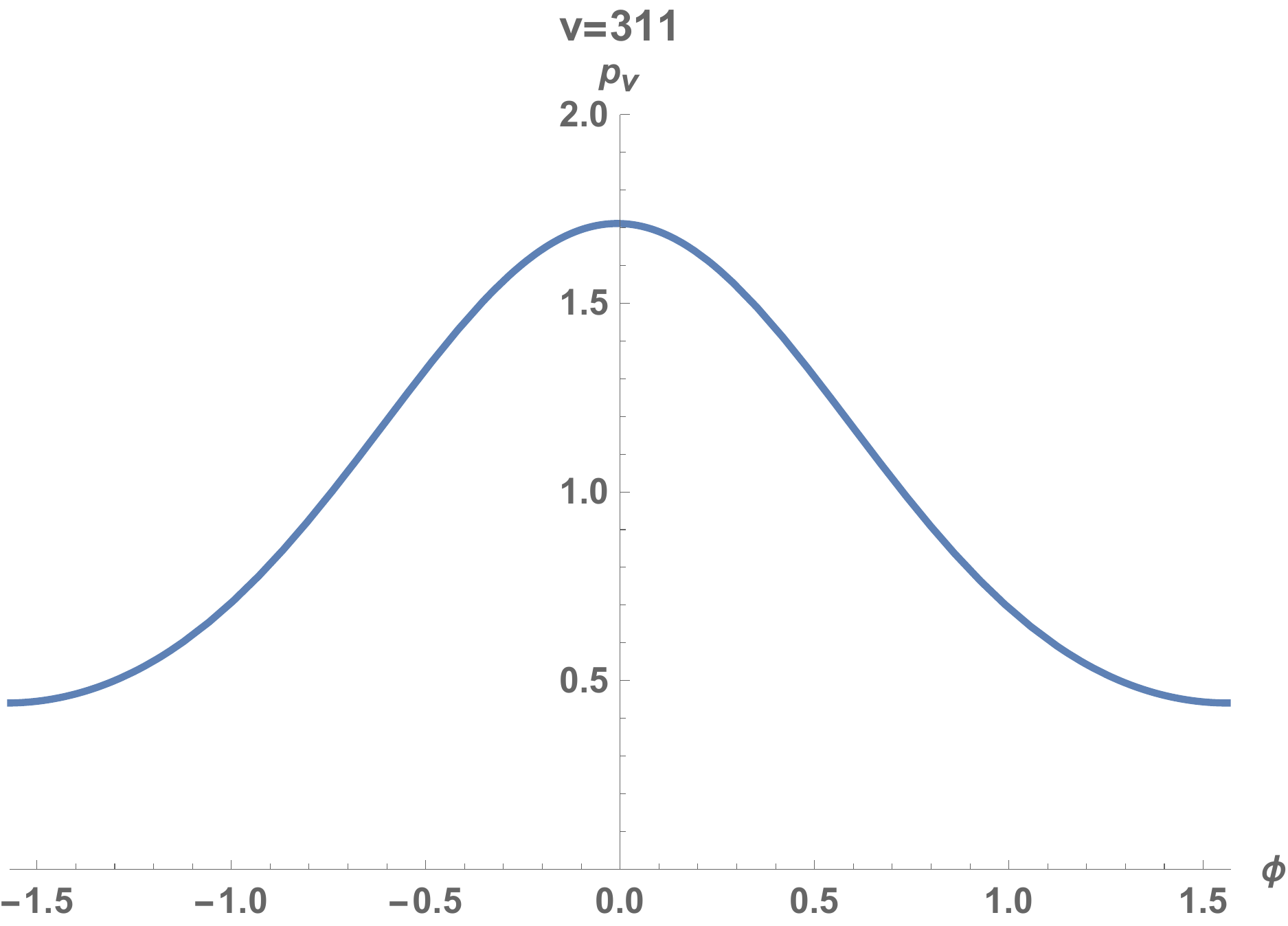}
 \end{center}
 \vspace{-5mm}
 \caption{Dynamical evolution of a perturbed Einstein black ring near $y=-1/R$, with $m=2$ and $m/R=0.98$. }
 \label{fig:nonlineaevomR098}
\end{figure}
\begin{figure}
 \begin{center}
  \includegraphics[width=34mm,angle=0]{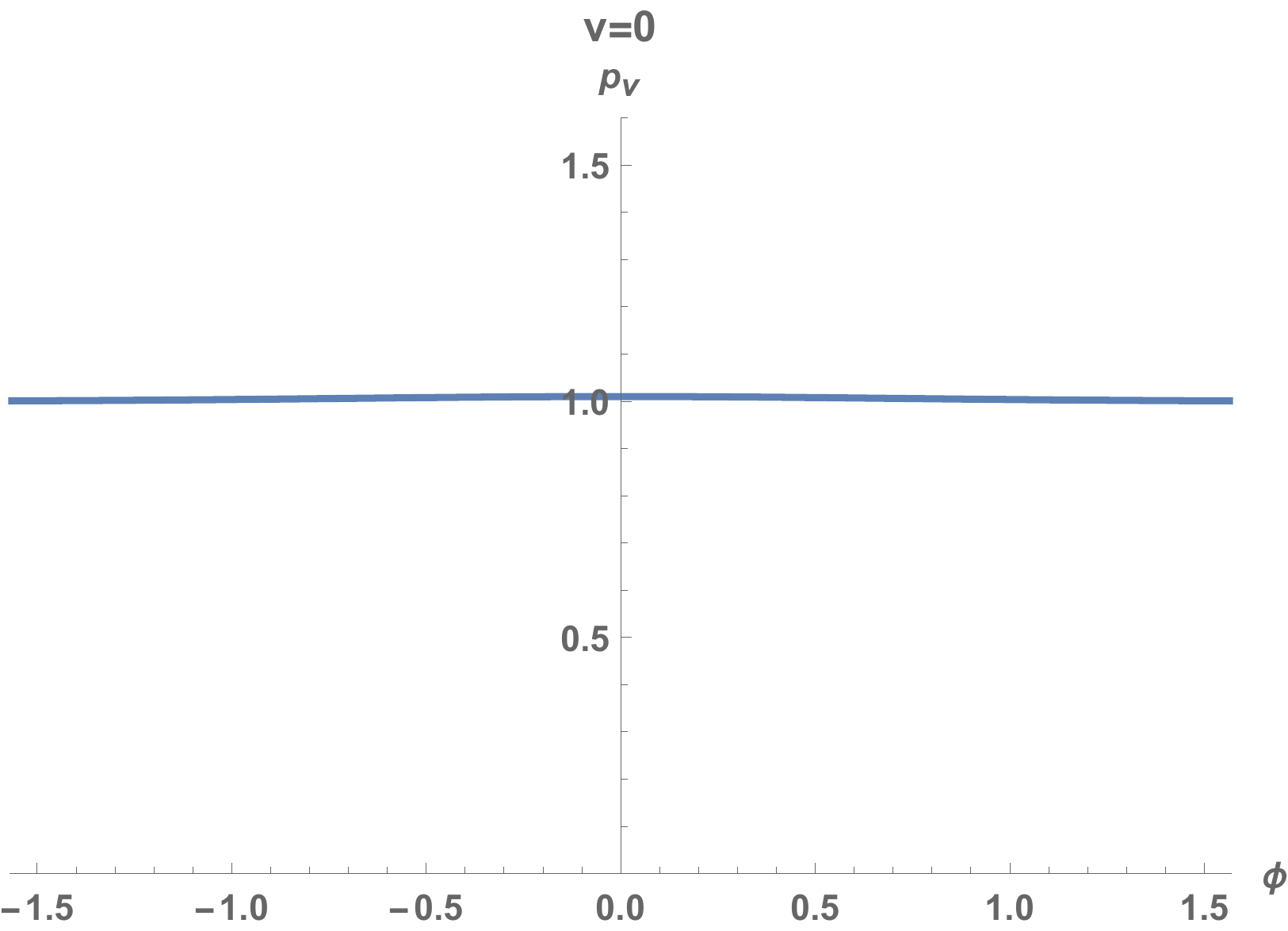}
 \hspace{1mm}
  \includegraphics[width=34mm,angle=0]{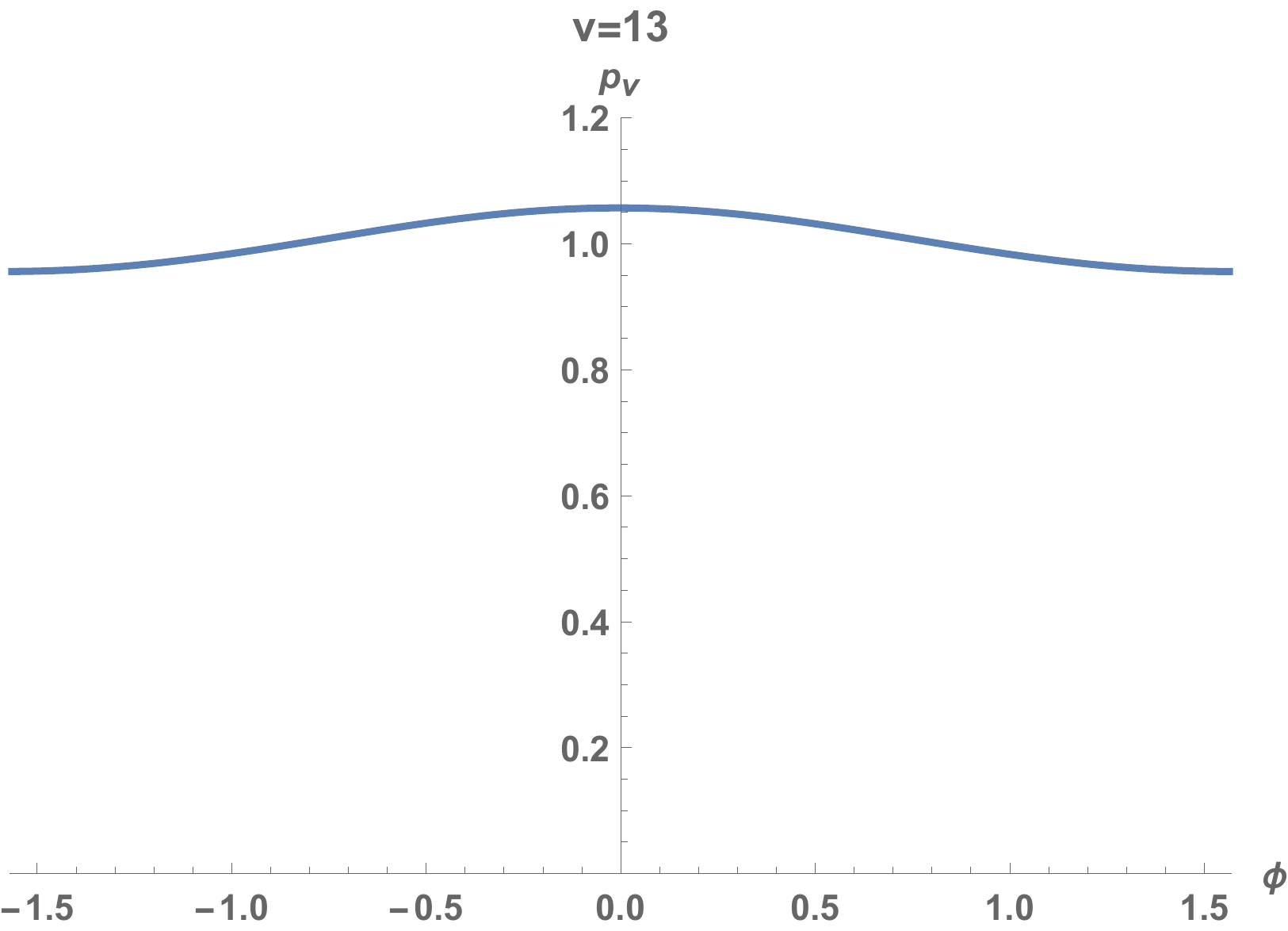}
   \hspace{1mm}
  \includegraphics[width=34mm,angle=0]{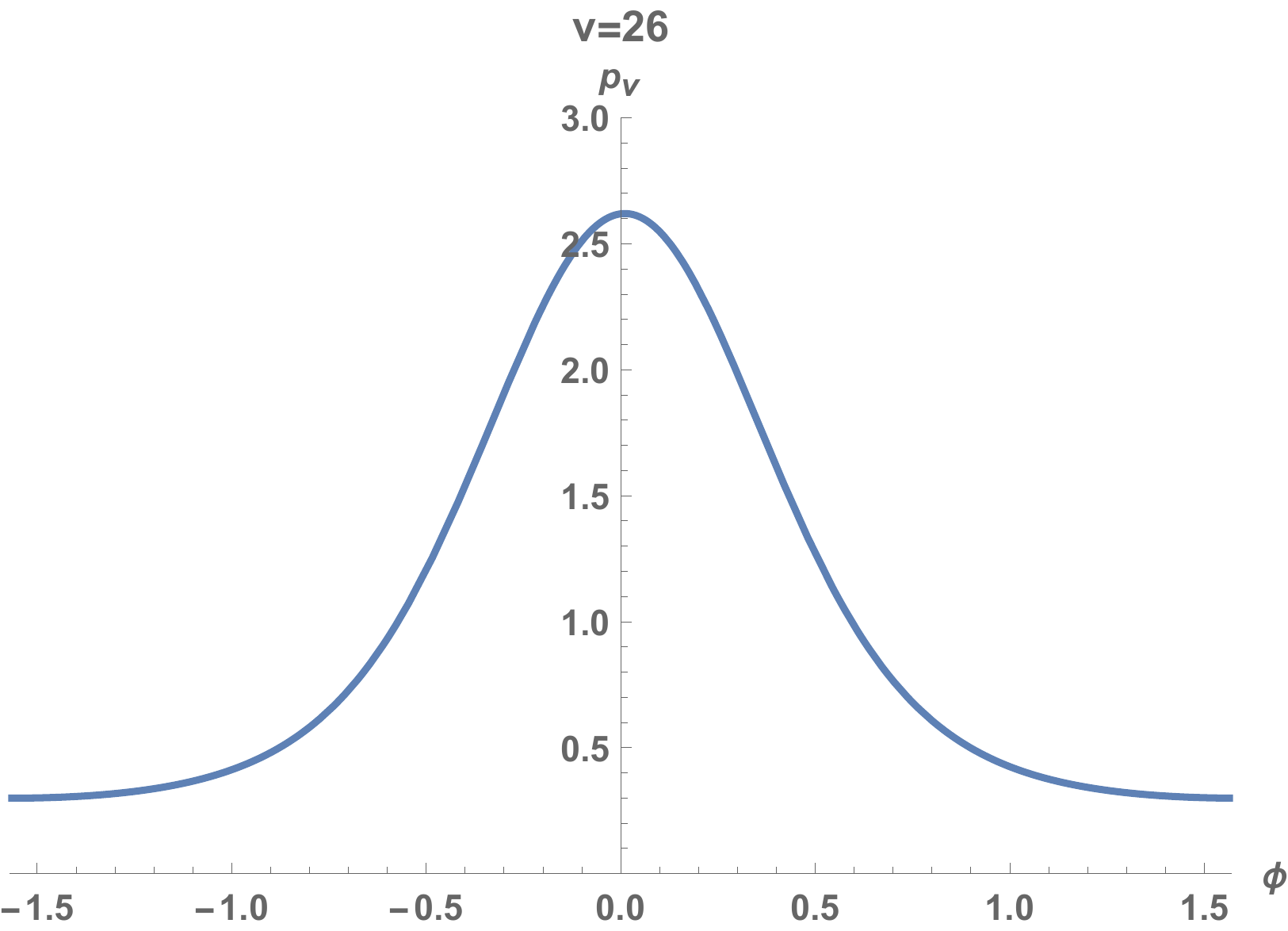}
  \hspace{1mm}
  \includegraphics[width=34mm,angle=0]{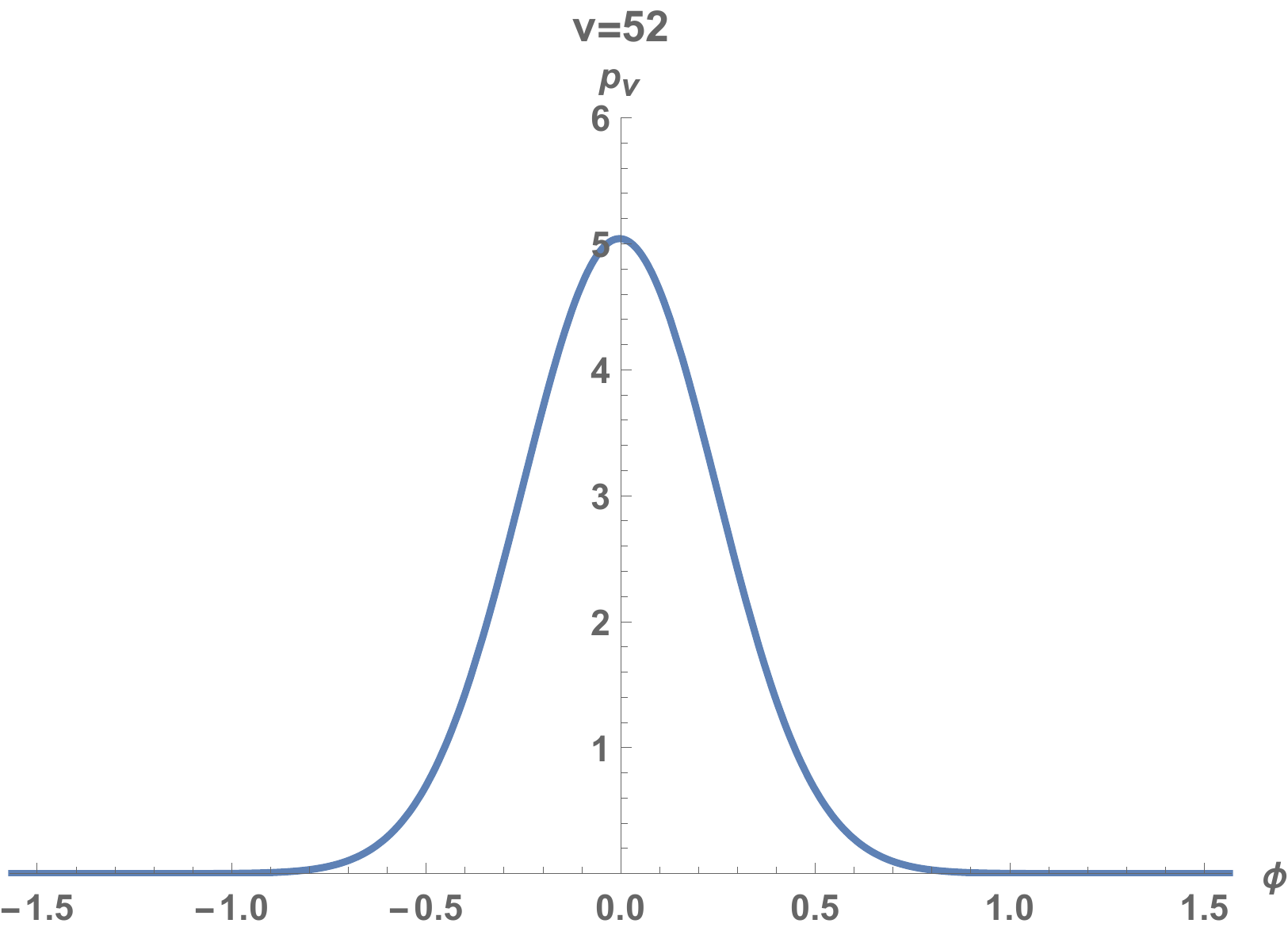}
 \end{center}
 \vspace{-5mm}
 \caption{Dynamical evolution of a perturbed Einstein black ring near $y=-1/R$, with $m=2$ and  $m/R=0.5$. }
 \label{fig:nonlineaevomR05}
\end{figure}
\begin{figure}
 \begin{center}
  \includegraphics[width=44mm,angle=0]{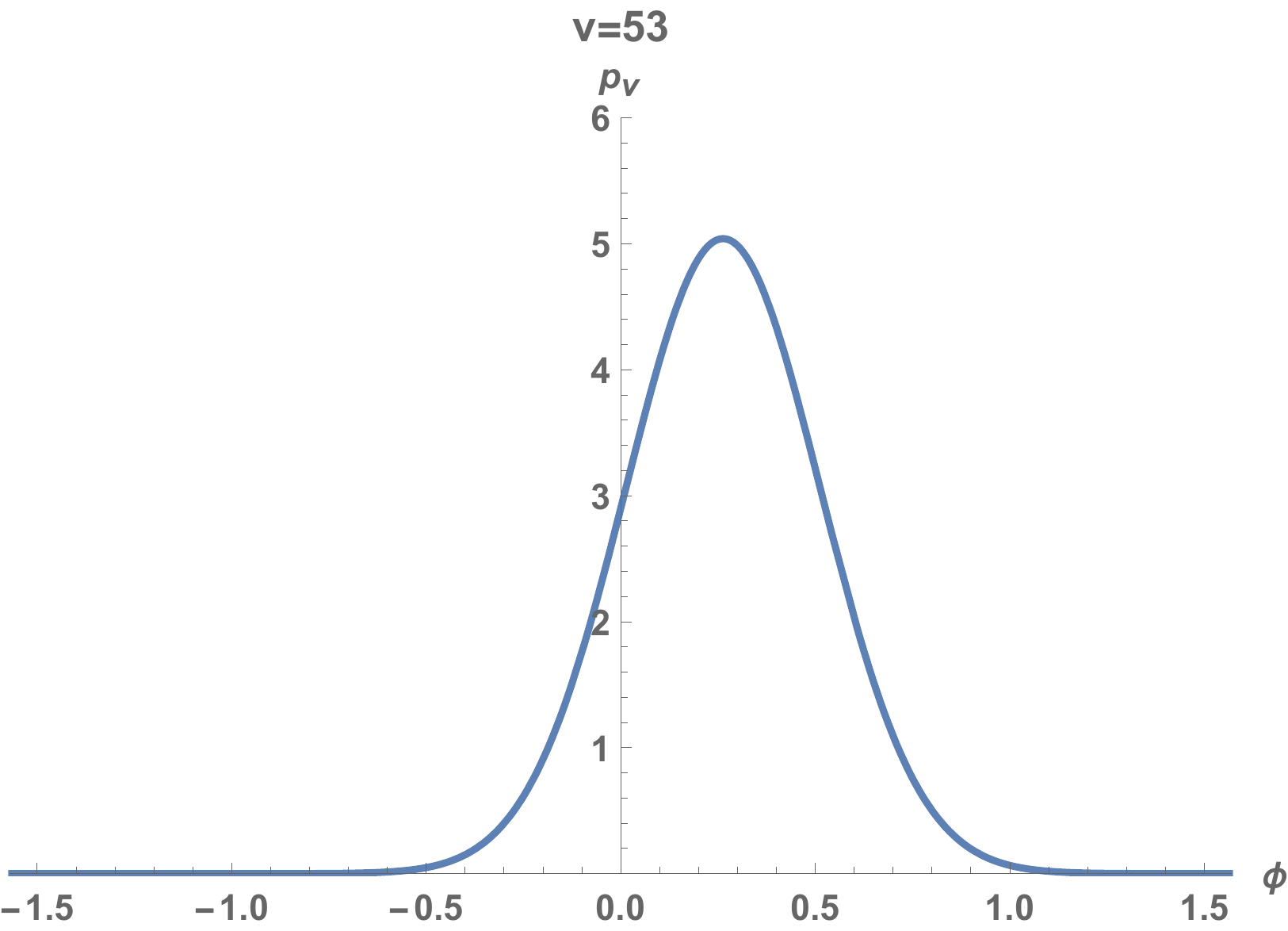}
 \hspace{1mm}
  \includegraphics[width=44mm,angle=0]{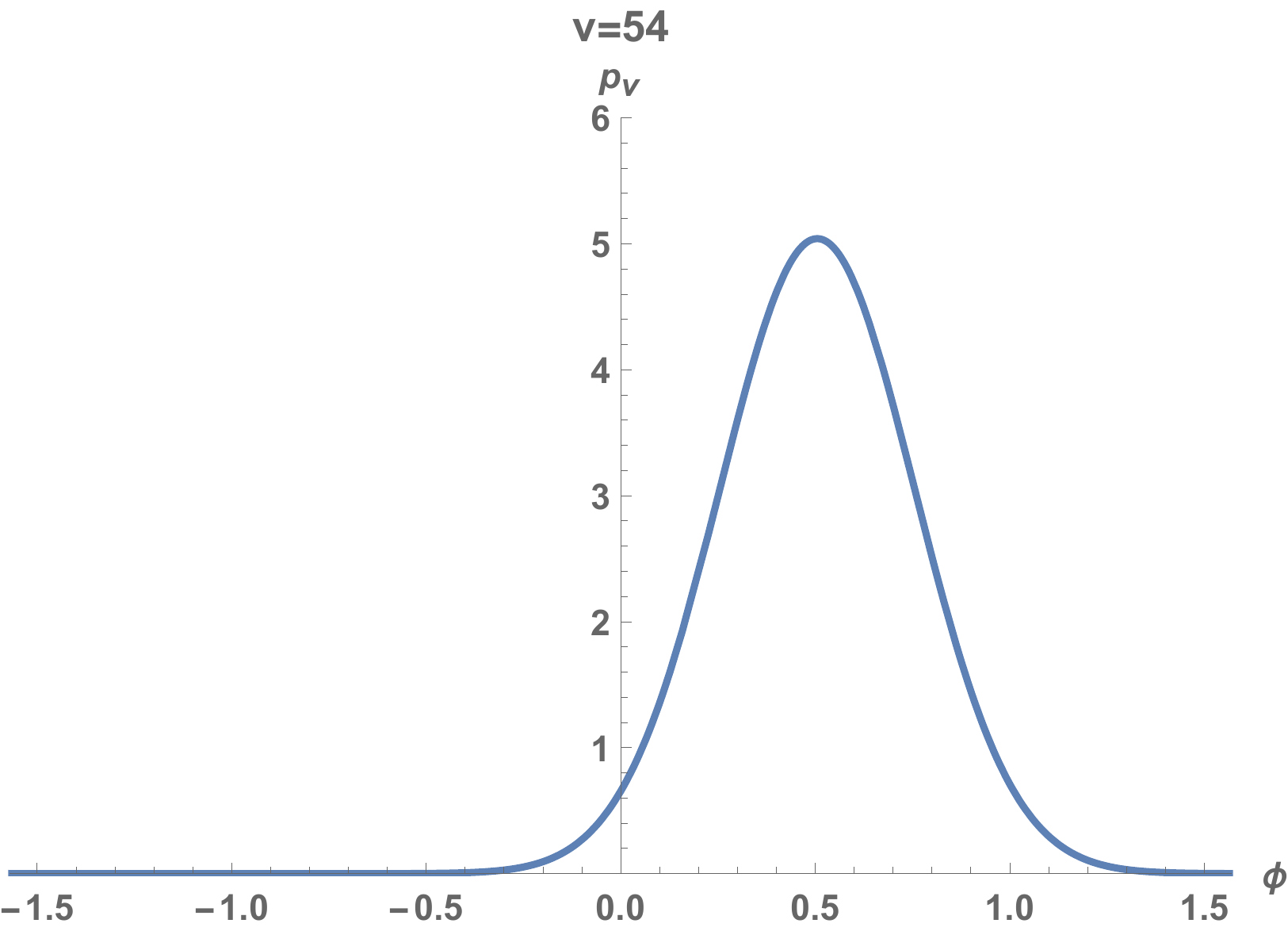}
   \hspace{1mm}
  \includegraphics[width=44mm,angle=0]{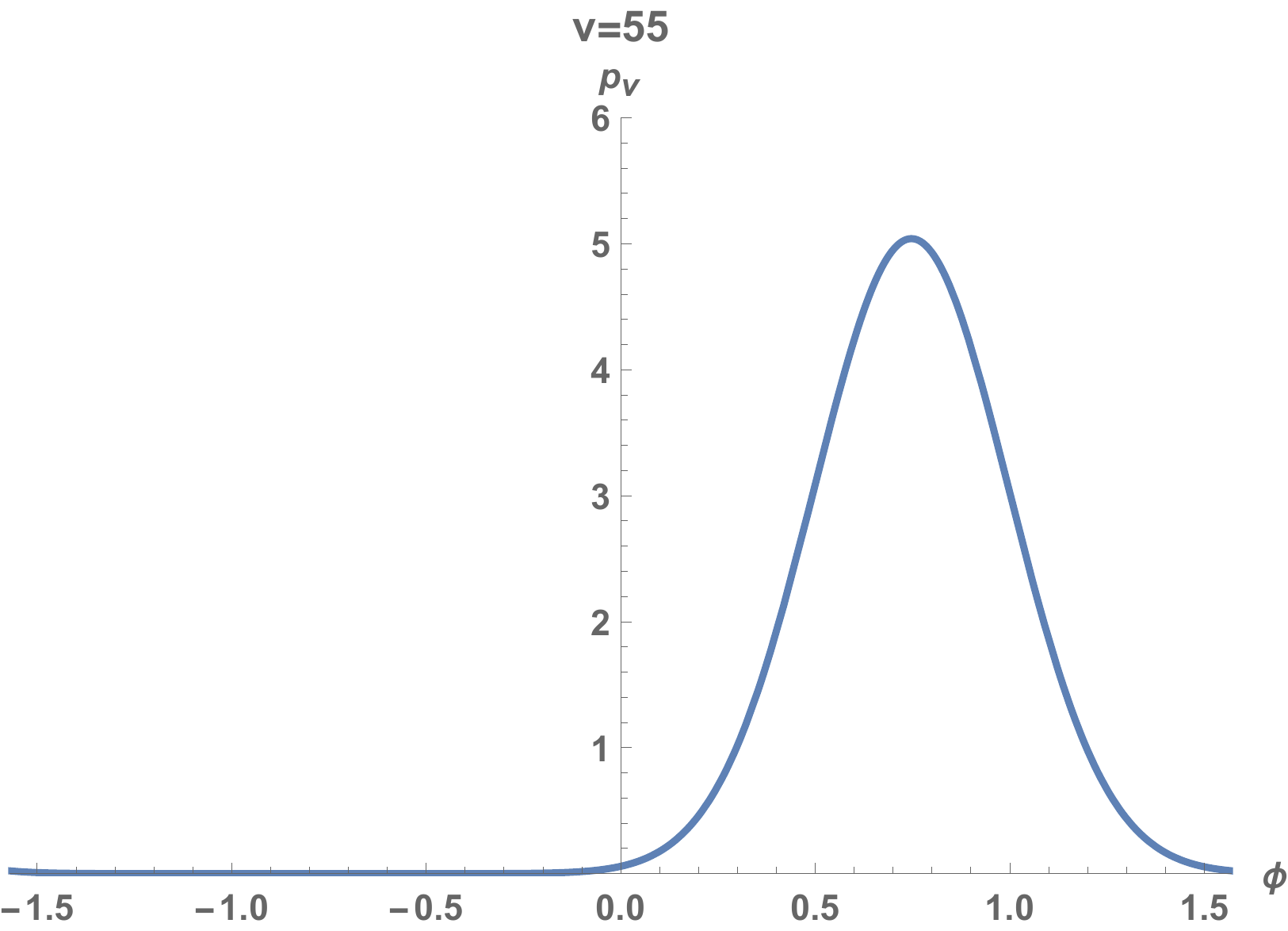}
 \end{center}
 \vspace{-5mm}
 \caption{The subsequent behavior of the evolution of a perturbed Einstein black ring near $y=-1/R$, with $m=2$ and  $m/R=0.5$. }
 \label{fig:nonlineaevoR05subsequent}
\end{figure}

For a given $R$ and $m$, we introduce a small perturbation of the uniform black ring, $p_v(0,\phi)=1+\delta p_{v}$, $p_\phi(0,\phi)=\frac{R^2}{\sqrt{R^2-1}}(1+\delta p_{v})$.
We find that when $R<m$, the perturbation quickly dissipates and the black ring becomes uniform along the $\phi$ direction,
this is consistent with the perturbation analysis in previous subsection that the linear modes are stable with $R<m$. On the contrary,
when $R>m$, the initial deformation grows fast, eventually the black ring settles down at a stable configuration that approximates well the non-uniform black ring obtained
as the stationary solution of the equations (\ref{Eq1:specialcase}) and (\ref{Eq2:specialcase}).
In Fig. \ref{fig:nonlineaevomR098} and Fig. \ref{fig:nonlineaevomR05} we show two typical results of the simulation. Since at $y=-1/R$ uniform black ring has $p_v=1$, thus by plotting $p_v(v,\phi)$
 the non-uniformity can be clearly demonstrated.
 Fig. \ref{fig:nonlineaevomR098} shows the evolution of a not-too-thin black ring with $m=2$ and $R=2.04$. Since the growth rate is small the evolution costs much time to reach the final stable state. Similar to the Einstein black string \cite{Emparan1506}, we find the final profile of the evolution is approximately cosinoidal.
 For a thinner black ring with $m=2$ and $R=4$, the evolution is faster and the final profile is very approximately gaussian, also similar to the Einstein black string.

 However, the distinction between the evolution of the black ring and that of the black string is apparent.
 From Fig. \ref{fig:nonlineaevoR05subsequent} we can see that when the evolution is stable, the amplitude of the  large blob along $\phi$ direction stops growing, however, along
$\phi$ direction the  large blob is not static but moves with time this is
 due to the existence of the angular momentum along that direction. Correspondingly, for slowly boosted black string, due to the existence of the angular momentum
 along the string direction we observe the similar phenomena.

\subsubsection{EGB black ring}
\begin{figure}
 \begin{center}
  \includegraphics[width=44mm,angle=0]{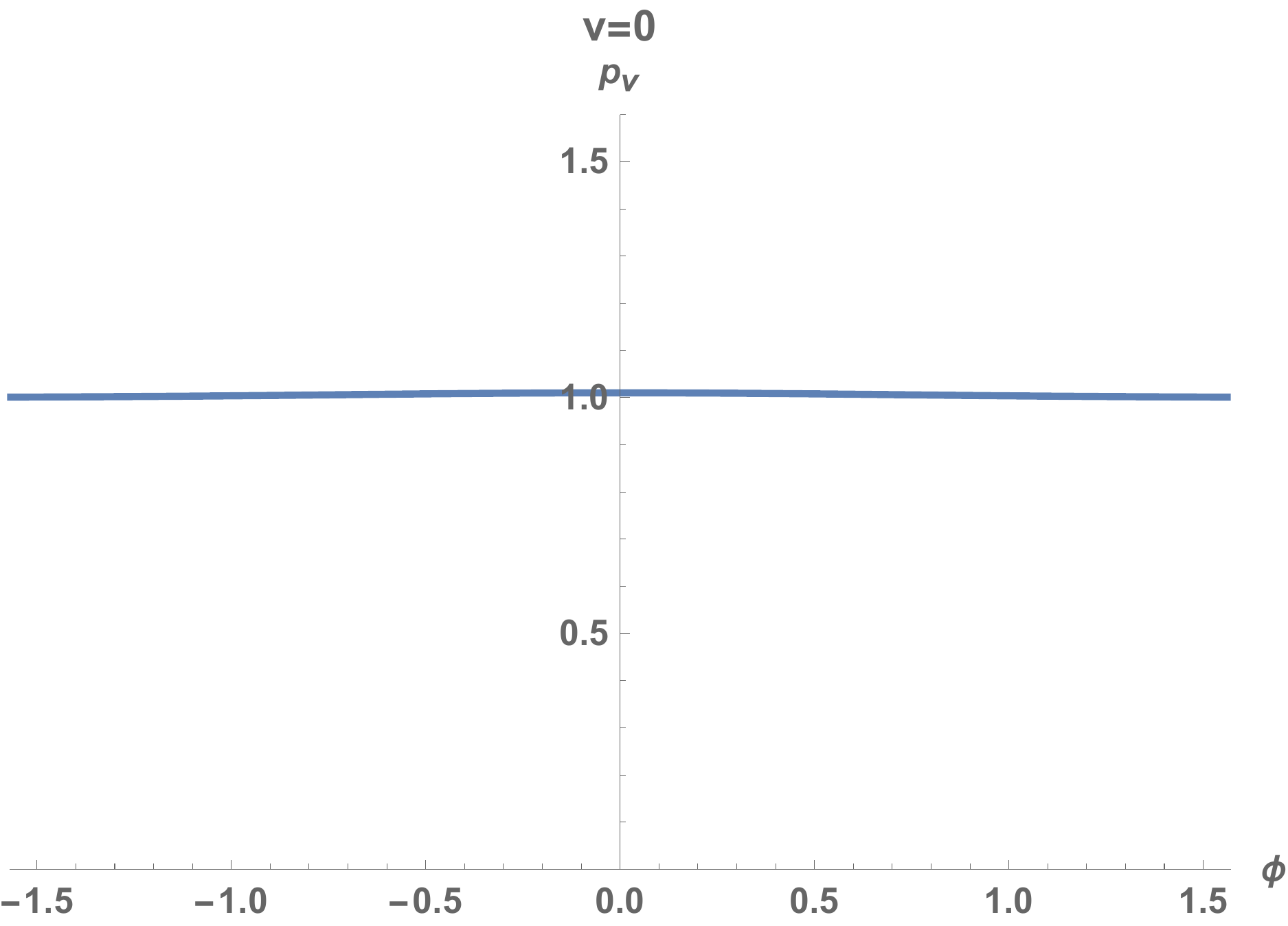}
 \hspace{1mm}
  \includegraphics[width=44mm,angle=0]{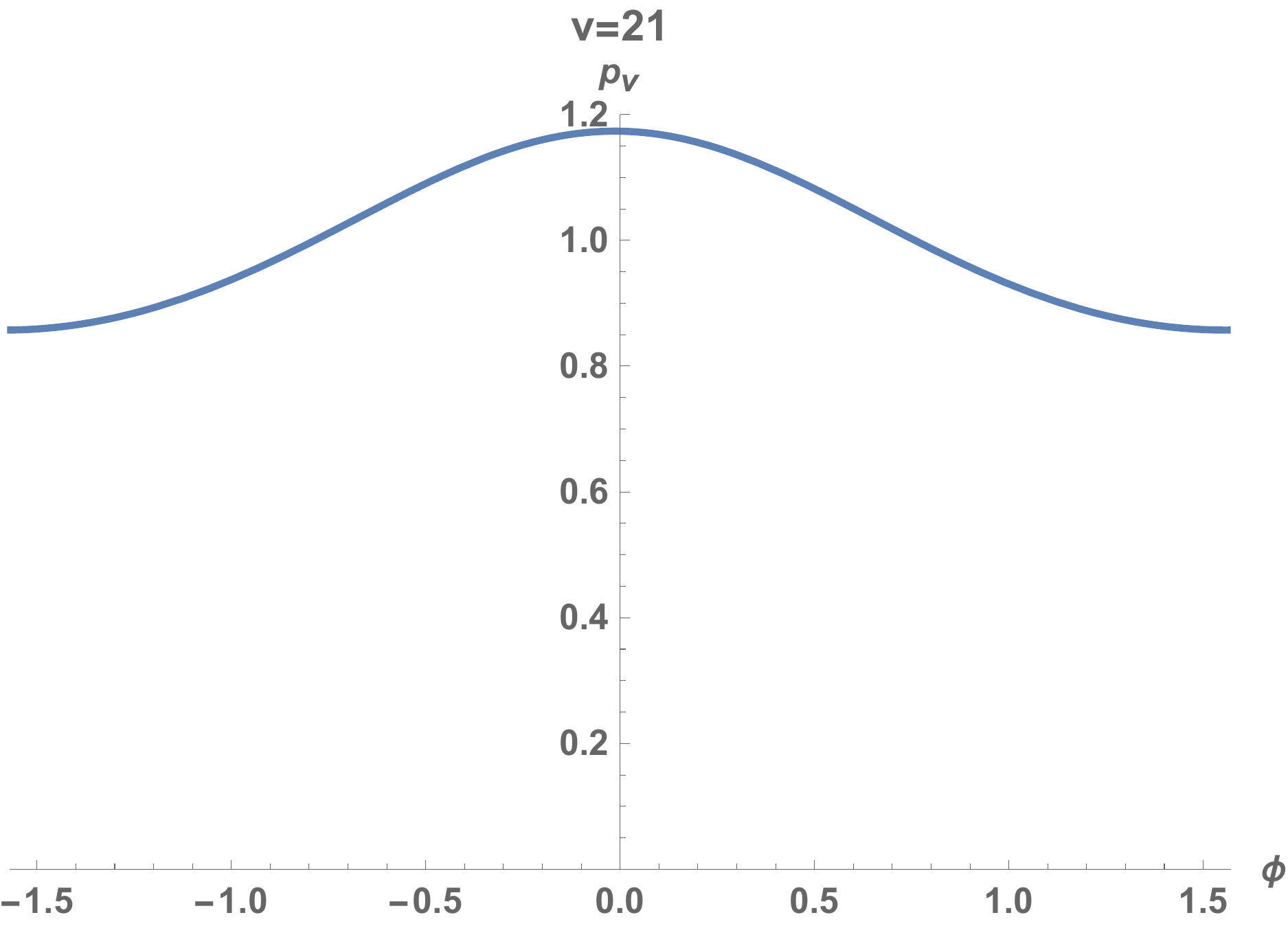}
  \hspace{1mm}
  \includegraphics[width=44mm,angle=0]{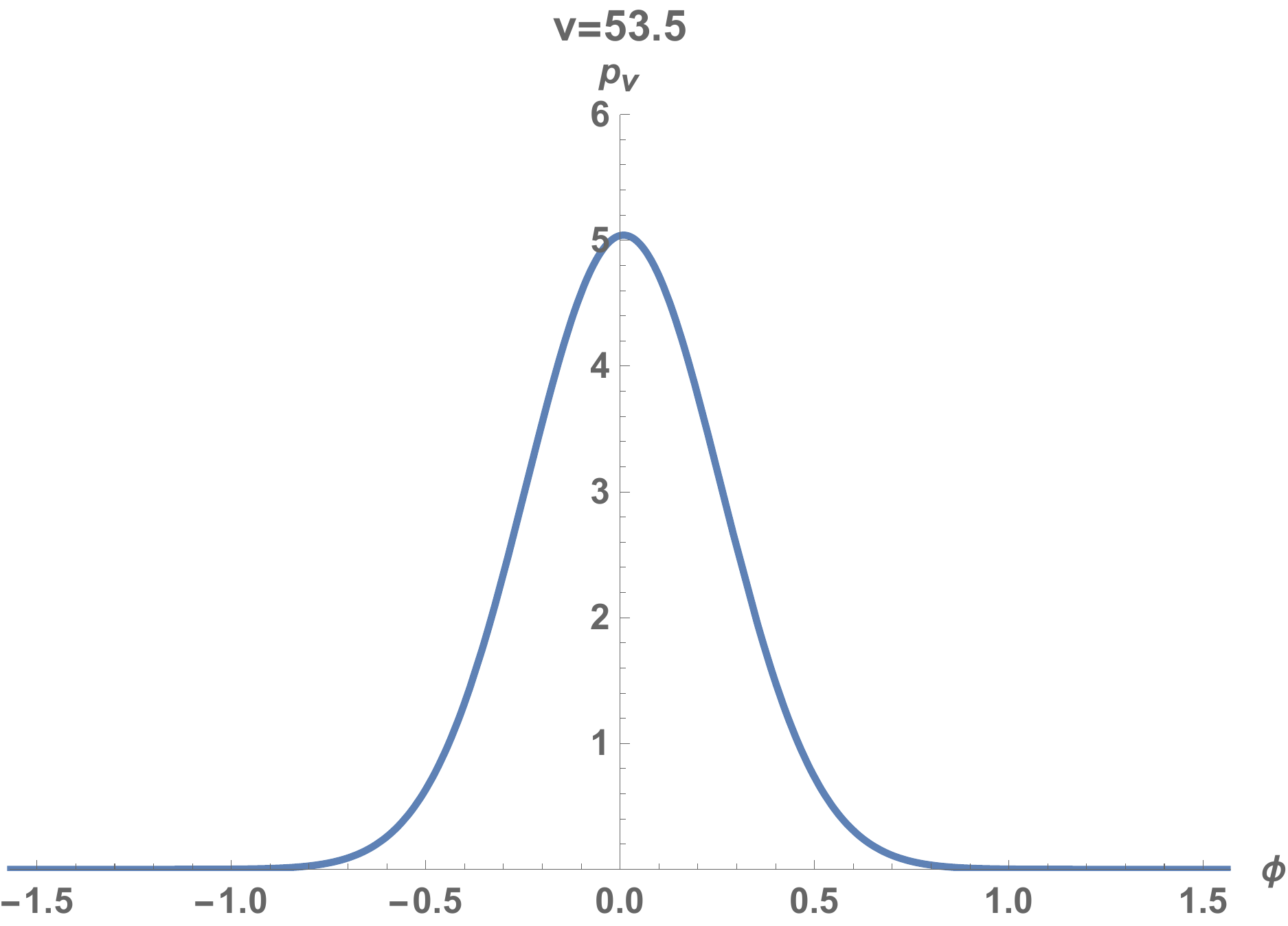}
 \end{center}
 \vspace{-5mm}
 \caption{Dynamical evolution of a perturbed EGB black ring near $y=-1/R$, with $m=2$, $m/R=0.5$ and $\tilde{\alpha}=2$. }
 \label{fig:nonlineaevomR05EGB}
\end{figure}

The effective equations for the EGB black ring  are lengthy and will not be given explicitly. We find that near
$y=-1/R$, $p_z$ is decoupled from the other two of the effective equations as well, then by numerically solving these two equations the non-linear evolution of the
EGB black ring can be demonstrated. Basically, the dynamical evolution of the unstable EGB black ring is very
similar to that of the Einstein black ring. That is, when $R<m$ the initial perturbation quickly dissipates, and when $R>m$ the perturbation grows fast and the black
ring settles down at the non-uniform black ring at late time. Fig. \ref{fig:nonlineaevomR05EGB} shows the evolution of a thin EGB black ring with $m=2$, $R=4$ and $\tilde{\alpha}=2$.

According to the discussion of the EGB black string \cite{Chen1707},
we know that the GB term affects the evolution rate of the black string which is consistent with the perturbation analysis.
 For the EGB black ring, from previous discussion the GB term affects both the angular momentum and the growth rate of the
 unstable mode, thus the presence of the GB term does not essentially change the above results.
 We find that when $\tilde{\alpha}$ is smaller than a critical value, the time that the solution needed to reach the stable state decreases with $\tilde{\alpha}$, and
 when $\tilde{\alpha}$ is larger than the critical value then the time needed to reach the stable state increases with $\tilde{\alpha}$, which
 is consistent with the perturbation analysis. The critical value of $\tilde{\alpha}$ is close to $\frac{R^2}{\sqrt{2}(R^2-1)}$.

 \section{Other slowly rotating black objects in EGB theory}\label{section5}
The large $D$ effective equations (\ref{Effeq1}), (\ref{Effeq2}) and (\ref{Effeq3}) derived in section \ref{section2} can actually describe other slowly rotating black objects, including the black holes and the black strings. In this section, we discuss these two cases briefly. The essential point in the construction is to determine the  functions $G(z)$ and $H(z)$ in different cases.

\subsection{Slowly Rotating Black Holes}
To discuss the slowly rotating black hole, we need to consider the metric of $D=n+4$ dimensional flat spacetime  in spherical coordinates
\be
ds^2=-dt^2+dr^2+r^2(dz^2+\mathrm{sin}^2\,z\,d\Phi^2+\mathrm{cos}^2\,z\,d\Omega_n^2).
\ee
As before, the embedding of the leading order metric is obtained by taking $r=1$. Then we find
\be\label{HandG:blackhole}
H(z)=\mathrm{cos}\,z,\qquad G(z)=\mathrm{sin}\,z.
\ee
From (\ref{kappahat}) and (\ref{surfacegra}) we obtain
\be
\hat{\kappa}=\frac{1}{2},\qquad \kappa=\frac{n}{2}\frac{1+\tilde{\alpha}}{1+2\tilde{\alpha}}.
\ee
As the surface gravity should be positive, we must have
\be
\tilde{\alpha}>-\frac{1}{2}.
\ee

\paragraph{Stationary solution} Directly from (\ref{staionarysol:pv}) and (\ref{staionarysol:pphi}) we obtain
\be
 p_z(z)=p_v'(z),\qquad p_\phi(z)= \hat{a}\, \sin^2z\, p_v(z),
 \ee
 where we have introduced $\hat{a}=\hat{\Omega}_H$. Furthermore $p_v(z)=e^{P(z)}$ can be determined by
 (\ref{stationarysol:Py}). After substituting (\ref{HandG:blackhole}) into (\ref{stationarysol:Py}) we have
 \be
 P''(z)+\tan z P'(z)-\frac{(1+\tilde{\alpha})(1+2\tilde{\alpha})\hat{a}^2\cos^2z }{1+\tilde{\alpha}+2\tilde{\alpha}^2}=0.
 \ee
 The solution of this equation is
 \be
 P(z)=p_0+d_0 \sin z-\frac{(1+\tilde{\alpha})(1+2\tilde{\alpha}) }{2(1+\tilde{\alpha}+2\tilde{\alpha}^2)}\hat{a}^2\cos^2z.
 \ee
 Similar to the case in \cite{Tanabe1510,Chen1702}, here the integration constants $p_0$ and $d_0$ describe trivial deformations of the solution,
 so we set them to be zero. Thus we completely obtain the stationary solution of the effective equations.
The metric of the solution is of the following form
 \bea\label{GB:slowlyrotatingbh}
 ds^2&=&-\Big(1+\frac{1-\Sigma}{2\tilde{\alpha}} \ \Big)dv^2+2dvdr+a\sin^2z\,\frac{1-\Sigma}{\tilde{\alpha}} \ dvd\Phi\nonumber\\
&&+ \frac{p_v'(z)}{n\, p_v(z)}\frac{1-\Sigma}{\tilde{\alpha}}dvdz+r^2\sin^2z\Big(1-a^2 \sin^2z\frac{1-\Sigma}{2\tilde{\alpha}}\Big)d\Phi^2\nonumber\\
&&+\frac{2r^2}{n}a\,p_v'(z)\sin^2z\frac{1-\Sigma}{2\tilde{\alpha}}dzd\Phi+r^2dz^2+ r^2\cos^2z\, d\Omega_n^2,
 \eea
 where we have used $\Sigma$ defined in (\ref{Sigmavzphi}) to simplify this expression. This metric describes the $D=n+4$ dimensional slowly rotating black holes in the EGB gravity, with $a=\hat{a}/\sqrt{n}$. When $\hat{a}=0$,  the solution is reduced to
 the spherically symmetric GB black hole \cite{Boulware1985,Wheeler1986}.
 Despite the fact that we do not have the exact form of rotating black holes in the EGB gravity, the slowly rotating  black holes with linear order of
  $a$ appearing in the metric has been obtained \cite{Kim0711}. It is easy to see that at the leading order of the $1/n$ expansion and
  up to the linear order of $a$ , the solution we obtained above (\ref{GB:slowlyrotatingbh}) simply reproduces the result of \cite{Kim0711}.
  This strongly supports that we have obtained the correct slowly rotating black hole solution in the EGB gravity with
  $a=\mc O(1/\sqrt{n})$.

\begin{figure}[t]
 \begin{center}
  \includegraphics[width=65mm,angle=0]{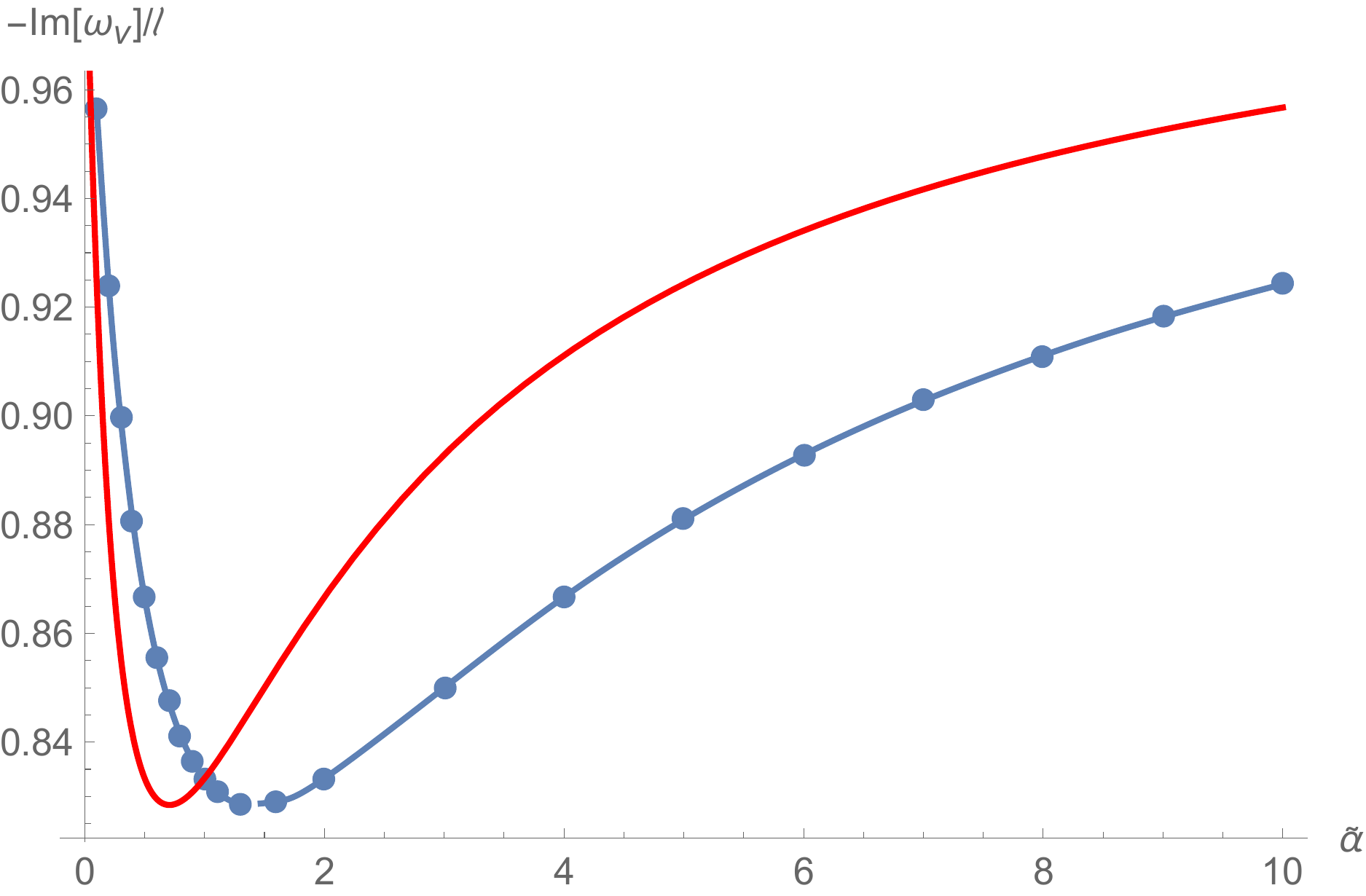}
 \end{center}
 \vspace{-5mm}
 \caption { The vector-type QNMs of the slowly rotating black hole in EGB theory. The dotted line is the result obtained by numerically solving the
 master equation. The solid line is the numerical evaluation of (\ref{vectortypeQNM}).  }
 \label{fig:vectortype}
\end{figure}

 \paragraph{Quasinormal modes}Now let us investigate the QNMs of the slowly rotating black hole in the EGB gravity. We first consider the following perturbations  around the aforementioned stationary solution
\bea
p_v(v,z,\phi)&=&e^{-\frac{(1+\tilde{\alpha})(1+2\tilde{\alpha})\hat{a}^2\cos^2z }{2(1+\tilde{\alpha}+2\tilde{\alpha}^2)}}\big(1+\epsilon F_v(z)e^{-i\omega v}e^{im\phi} \big),\\
p_z(v,z,\phi)&=&\frac{(1+\tilde{\alpha})(1+2\tilde{\alpha})\hat{a}^2\sin z\cos z }{(1+\tilde{\alpha}+2\tilde{\alpha}^2)}\big(1+\epsilon F_z(z)e^{-i\omega v}e^{im\phi} \big),\\
p_\phi(v,z,\phi)&=&\hat{a}\sin^2\,z\,e^{-\frac{(1+\tilde{\alpha})(1+2\tilde{\alpha})\hat{a}^2\cos^2z }{2(1+\tilde{\alpha}+2\tilde{\alpha}^2)}}\big(1+\epsilon F_\phi(z)e^{-i\omega v}e^{im\phi} \big).
\eea
Then plugging these perturbations into the effective equations, and taking in account that at $z=0$, $F_v(z)\propto z^\ell$ \cite{Emparan1402}, we can obtain the frequencies of the QNMs. For simplicity here we only consider the case  $m=\mc O(1/n)$, since other cases are more
involved \cite{Tanabe1510}. Thus we may introduce another  $\mc O(1)$ quantity by
 \be
\bar{m}=nm.
\ee
 Then the QNM frequencies are characterized by $\ell$, $\bar{m}$ and $\hat{a}$.
\paragraph{Vector-type gravitational perturbation} For the slowly rotating black hole, it is possible to discuss the vector-type gravitational perturbation. We obtain the QNM frequency
\be\label{vectortypeQNM}
\omega_v=-i\ell\frac{1+2\tilde{\alpha}+2\tilde{\alpha}^2}{(1+\tilde{\alpha})(1+2\tilde{\alpha})},
\ee
which suggests that the perturbation is stable in EGB theory. Both in the small and the large $\tilde{\alpha}$ limits, we obtain
\be
\omega_v=-i\ell,
\ee
which agrees exactly with the ones obtained  by directly computing from the master equations of the perturbations \cite{Chen1511}. For a generic GB parameter  $\tilde{\alpha}=\mc O(1)$, the vector-type QNM is only obtained by  numerically  solving
the mater equations. As shown in Fig. \ref{fig:vectortype} we  can see that the result matches well with the numerical value of (\ref{vectortypeQNM}), .

\paragraph{Scalar-type gravitational perturbation} For the scalar-type gravitational perturbation, we have
\bea
\omega_0&=&-i(\ell-2)\frac{1+2\tilde{\alpha}+2\tilde{\alpha}^2}{(1+\tilde{\alpha})(1+2\tilde{\alpha})},\\
\omega_{\pm}&=&\frac{-i(\ell-1)(1+2\tilde{\alpha}+2\tilde{\alpha}^2)\pm \sqrt{(\ell-1)\Big[(1+2\tilde{\alpha}+2\tilde{\alpha}^2)^2-\ell \tilde{\alpha}^2 \Big]}}{(1+\tilde{\alpha})(1+2\tilde{\alpha})}.
\eea
Note that the modes with the frequency $\omega_v$ and $\omega_0$ correspond to the vector-type gravitational perturbations of the GB black hole with $\ell=\ell_v-1$ and $\ell_v+1$ respectively, where $\ell_v$ is the angular momentum of the vector harmonics on $S^{D-2}$.
The modes with $\omega_{\pm}$ are just the QNMs of the GB black hole \cite{Chen1703}. Since we are considering a slowly rotating black hole with $m=\mc O(1/n)$, at the leading order  the dependence of the perturbation on $\phi$ disappears and the QNMs are captured by the ones of the  GB black hole with one lower dimensions. At the next order of the $1/n$ expansion, like the case in the Einstein gravity \cite{Tanabe1510} $\bar{m}$ will appear in the QNMs.

\subsection{Slowly boosted black string}
To discuss the black string, we consider the metric of $D=n+4$ dimensional spacetime  with one compact direction in the spherical coordinates
\be
ds^2=-dt^2+d\Phi^2+dr^2+r^2(dz^2+\sin^2 z\, d\Omega^2_n),
\ee
where $\Phi$ is the coordinate of the compact direction. The embedding of $r=1$ of the leading order metric into the above background gives the  following identifications
\be
G(z)=1,\quad H(z)=\mathrm{sin}\,z.
\ee
Since the embedded solution has $S^1\times S^{n+1}$ horizon topology, the effective equations describe the non-linear dynamics of the slowly boosted black string solution in the EGB gravity.
For stationary solution we have
\be
p_v(z)=e^{P(z)},\quad p_\phi=p_\phi(z),\quad p_z=p_z(z).
\ee
\paragraph{Stationary solution} From (\ref{staionarysol:pv}) and (\ref{staionarysol:pphi}) we have
\be
p_z(z)=m'(z),\quad p_\phi(z)=\hat{\sigma} m(z).
\ee
Here $\hat{\sigma} $ is an integration constant  describing the boost effect of the black string. Furthermore, the equation for $P(z)$ is given by (\ref{stationarysol:Py}) which now becomes
 \be
 P''(z)-\mathrm{cot}\,z\, P'(z)=0,
 \ee
which can be solved easily and yields
\be
p_v(z)=e^{p_0+p_1 \mathrm{cos}\,z}.
\ee
 The integration constants $p_0$ and $p_1$ are trivial deformations of the solution  so we may  set them to be zero.
Then the leading order metric of the slowly boosted EGB black string is given by
\bea\label{boostedEGBblackstring}
 ds^2&=&-\Big(1+\frac{1-\Sigma}{2\tilde{\alpha}} \ \Big)dv^2+2dvdr-\frac{\hat{\sigma}}{\sqrt{n}}\,\frac{\Sigma-1}{\tilde{\alpha}} \ dvd\Phi\nonumber\\
&&+\Bigg[1+\frac{1}{n}\Bigg(\hat{\sigma}^2\frac{\Sigma-1}{2\tilde{\alpha}}-\frac{2\ln\frac{1+\Sigma}{2}+\frac{\pi}{2}-2\arctan \Sigma-\frac{1}{(1+2\tilde{\alpha})}\ln\frac{1+\Sigma^2}{2}}{\tilde{\alpha}+1}\Bigg)\Bigg]d\Phi^2\nonumber\\
&&+r^2dz^2+ r^2\sin^2z\, d\Omega_n^2.
 \eea
The metric is related to the uniform EGB black string constructed in \cite{Chen1707}  by a boost transformation with the boost velocity
 \be\label{boostedEGBblackstring:boost}
\mathrm{sinh}\,\gamma=\frac{\hat{\sigma}}{\sqrt{n}}.
\ee
Comparing (\ref{boostedEGBblackstring}) with (\ref{largeRlimitmetric}), we see that when the boost velocity of the slowly boosted black string takes a specific value, i.e. $\gamma=\beta$ with $\beta$ being defined by (\ref{largeRlimitboost}), then the large radius limit of the EGB black ring is reproduced.

\paragraph{Non-uniform black string} Another stationary solution of the effective equations is the non-uniform black string which is believed to be the end point of the evolution of the large $D$ EGB black string, as showed in \cite{Chen1707}. The non-uniformity means that the solution is inhomogeneous along the $\Phi$ direction. In this case, there is only one Killing vector $\partial_v$. Moreover, we do not need the $z$ dependence anymore, so $p_z$ can be set to zero. Then we have
\be
p_v=p_v(\phi), \quad p_\phi=p_\phi(\phi).
\ee
From the effective equations we find
\be
\quad  p_\phi=p_v'(\phi),
\ee
and
\be
p_v'''-\frac{2p_v'\,p_v''}{p_v}+p_v'+\frac{(1+\tilde{\alpha})(1+2\tilde{\alpha})}{1+\tilde{\alpha}+2\tilde{\alpha}^2}\frac{p_v'^3}{p_v^2}=0.
\ee
This equation can be solved numerically, as discussed in \cite{Chen1707}.
\paragraph{Quasinormal modes} Let us now investigate the QNMs of the slowly boosted black string we constructed above. We consider the perturbations around the stationary solution,
\bea
p_v(v,z,\phi)&=&1+\epsilon F_v(z)e^{-i\omega v+i\hat{k}\phi},\\
p_z(v,z,\phi)&=&\epsilon F_z(z)e^{-i\omega v+i\hat{k}\phi},\\
p_\phi(v,z,\phi)&=&\hat{\sigma}\big(1+\epsilon F_\phi(z)e^{-i\omega v+i\hat{k}\phi} \big).
\eea
As before, plugging these expressions into the effective equations and taking into account that at $\cos z=0$, $F_v(z)\propto \cos^\ell z$,
we obtain the QNM condition for the scalar-type gravitational perturbation as follows.
\bea\label{QNMcondition:boostEGBblackstring}
&&\omega^3+\omega^2\Big[-3\hat{\sigma}\hat{k}+\frac{i(-2+3\hat{k}^2+3\ell)(1+2\tilde{\alpha}+2\tilde{\alpha}^2)}{(1+2\tilde{\alpha})(1+\tilde{\alpha})}\Big]\nonumber\\
&&-\frac{\omega}{(1+2\tilde{\alpha})^2(1+\tilde{\alpha})^2}\Bigg[\Big(12 \tilde{\alpha}^4+24 \tilde{\alpha}^3+23 \tilde{\alpha}^2+12 \tilde{\alpha}+3\Big)\Big(\hat{k}^4+2\ell\hat{k}^2-\ell(\ell-1)\Big)\nonumber\\
&& -\hat{k}^2\Big(3 \hat{\sigma}^2(1+2\tilde{\alpha})^2(1+\tilde{\alpha})^2+20 \tilde{\alpha}^4+40 \tilde{\alpha}^3+39 \tilde{\alpha}^2+20\tilde{\alpha}-5\Big)\nonumber\\
&&+(6i\hat{k}^3+2i(3\ell-2)\hat{k})\hat{\sigma}\Big(1+5\tilde{\alpha} + 10\tilde{\alpha}^2 + 10\tilde{\alpha}^3 + 4\tilde{\alpha}^4\Big)\Bigg]\nonumber\\
&&-\frac{i}{(1+2\tilde{\alpha})^4(1+\tilde{\alpha})^4}\Bigg[(1 + 3 \tilde{\alpha} + 6 \tilde{\alpha}^2 + 6 \tilde{\alpha}^3 + 4 \tilde{\alpha}^4)(\hat{k}^6+(\ell -1) \ell ^2)+\nonumber\\
&&+\Big(3\hat{k}^4(\ell -1)+\hat{k}^2(3 \ell ^2-4 \ell +2)\Big)(2\tilde{\alpha}^2+\tilde{\alpha}+1) (2\tilde{\alpha}^2+2\tilde{\alpha}+1)
\nonumber\\
&&-\Big(3\hat{k}^4+(3\ell-2)\hat{k}^2\Big)\hat{\sigma}^2(2 \tilde{\alpha}^2+2\tilde{\alpha}+1)(2 \tilde{\alpha}^2+3 \tilde{\alpha}+1)\nonumber\\
&&+i\hat{\sigma}\Big(\hat{k}^5+2\ell\hat{k}^3+(\ell -1) \ell\hat{k}\Big) (12\tilde{\alpha}^4+24\tilde{\alpha}^3+23 \tilde{\alpha}^2+12 \tilde{\alpha}+3)\nonumber\\
&&-i\hat{k}^3 \Big(\hat{\sigma}^3(2\tilde{\alpha}^2+3 \tilde{\alpha}+1)^2+\hat{\sigma}(20\tilde{\alpha}^4+40\tilde{\alpha}^3+39\tilde{\alpha}^2+20\tilde{\alpha}+5)\Big)\Bigg]=0.
\eea
As we explained above, the slowly boosted EGB black string is related to the large radius limit of EGB black ring once
the boost velocity takes a specific value. The QNMs on both sides should be related by such relation as well.
Indeed, by taking  the large radius limit, identifying $\hat{m}=\hat{k}$ and using $\zeta=\hat{\sigma}$,
we find that (\ref{QNMcondition:EGBblackring}) is identical to (\ref{QNMcondition:boostEGBblackstring}).

For $\ell=0$, this cubic equation of $\omega$ is easy to solve,
\bea
\omega_0^{(\ell=0)}&=&\hat{k} \hat{\sigma} -\frac{i (2\tilde{\alpha}^2+2 \tilde{\alpha}+1)(\hat{k}^2-2)}{(\tilde{\alpha}+1) (2\tilde{\alpha}+1)},\\
\omega_{\pm}^{(\ell=0)}&=&\hat{k} \hat{\sigma} -i\hat{k}^2\frac{1+2\tilde{\alpha}+2\tilde{\alpha}^2}{(1+\tilde{\alpha})(1+2\tilde{\alpha})}\pm\frac{i \hat{k}\sqrt{1+4\tilde{\alpha}+(7+\hat{k}^2)\tilde{\alpha}^2+8\tilde{\alpha}^3+4\tilde{\alpha}^4}}{(1+\tilde{\alpha})(1+2\tilde{\alpha})}.
\eea
The mode $\omega_0^{(\ell=0)}$ should be discarded as it is a gauge mode.
The modes $\omega_{\pm}^{(\ell=0)}$ are physical and correspond to the QNMs of the S-wave sector of the scalar-type
gravitational perturbation of the slowly boosted black string. When $\hat{\sigma}=0$, the above result reproduces the one
of the EGB black string \cite{Chen1707}. The solution branch $\omega_{+}^{(\ell=0)}$ shows the GL instability \cite{Gregory9301052} for $\hat{k}<1$. It is easy to see that the critical wavenumber of the threshold mode is independent of the GB coefficient. Moreover, the
effect of the GB term on the instability is the same as the one without the boost transformation \cite{Chen1707}. That is, when $-1/2<\tilde{\alpha}<1/\sqrt{2}$, the
increase of $\tilde{\alpha}$ alleviates the instability but when $\tilde{\alpha}>1/\sqrt{2}$ the increase of $\tilde{\alpha}$ enhances the instability.

For $\hat{k}=0$, the quasinormal mode condition yields
\bea
\omega_0^{(\hat{k}=0)}&=&-i\ell\frac{1+2\tilde{\alpha}+2\tilde{\alpha}^2}{(1+\tilde{\alpha})(1+2\tilde{\alpha})},\\
\omega_{\pm}^{(\hat{k}=0)}&=&\frac{-i(\ell-1)(1+2\tilde{\alpha}+2\tilde{\alpha}^2)\pm \sqrt{(\ell-1)\Big[(1+2\tilde{\alpha}+2\tilde{\alpha}^2)^2-\ell \tilde{\alpha}^2 \Big]}}{(1+\tilde{\alpha})(1+2\tilde{\alpha})}.
\eea
Similar to the analysis on the  axisymmetric perturbations ($m=0$) of the EGB black ring,  these modes correspond to the decoupled vector and scalar modes of EGB black hole on $S^{D-2}$. For $\ell\neq 0$ and $\hat{k}\neq0$, no unstable mode is found. Therefore for the slowly boosted GB black string, similar to the case in the Einstein gravity \cite{Tanabe1510}, the instability only exists in the S-wave ($\ell=0$) sector.

\section{Summary}\label{section6}
In this paper, by using the large $D$ effective theory of the black hole, we studied the slowly rotating  black holes in the
Einstein-Gauss-Bonnet (EGB) theory. These black holes include the EGB black ring, the slowly rotating EGB black hole and the
slowly boosted EGB black string. According to the property of the EGB black string that the tension is small relative to the mass,
and inspired by the analysis of thin black ring via the blackfold method at large $D$, it is believed that the large $D$ EGB black
ring belongs to the class of slowly rotating black holes  with the horizon angular velocity being of $\mc O (1/\sqrt{D})$. Thus the construction
 of the EGB black ring is feasible.

Firstly, by integrating out the radial direction of the EGB equations, we obtained the effective equations for the slowly
rotating black holes. The EGB black ring solution was constructed analytically as the stationary solution of the effective
equations with the embedding into flat space in the ring coordinates. We found that the horizon velocity and
the angular momentum decrease with the GB coefficient when it is small, however, when the GB coefficient is large, the
the horizon velocity and the angular momentum increase with it. By performing the perturbation analysis of the effective equations we obtained the
 quasinormal modes of the EGB black ring. We found the GL-like instability for the non-axisymmetric perturbations when the EGB
  black ring is relatively thin. The effect of the GB term on the instability is the same as it has on the angular momentum of the
  EGB black ring. As byproducts, we constructed the  slowly rotating
EGB black hole and the slowly boosted EGB black string via different embeddings, and
studied their dynamical instability. In particular, we found that the large radius limit of the EGB black ring
is identical to the slowly boosted EGB black string with a critical boost velocity.

By numerically solving the effective equations of the black ring, we studied the non-linear evolution of the EGB black ring.
We found that at $y=-1/R$, $p_z$ is decoupled from the other two functions such that the effective equations can be solved numerically.
In this particular case, the behavior of the black ring in the non-linear regime is very similar to that of the black string. When the
black ring is not very thin, i.e. $m/R>1$, the initial perturbation quickly dissipates and the black ring becomes uniform. For thinner black ring,
i.e. $m/R<1$, the initial deformation grows fast, finally the solution settles down at a stable non-uniform black ring. Therefore the numerical
result gives a relatively strong evidence to support the conjecture given in \cite{Emparan1506}.  Due to the existence of
the angular momentum, the inhomogeneity  along the rotating direction of the black ring is not static but moves with time.
The presence of the GB term does not essentially change the behavior of the black ring in the non-linear regime. The effect of
the GB term is reflected in the time the unstable black ring needed to reach the final stable state. Similar to the effect on the
black string, we found that when
$\tilde{\alpha}$ is small, the time  to reach the stable state decreases with it, and when $\tilde{\alpha}$ is large,
 then the time increases with it, which is qualitatively consistent with the perturbation analysis.

 The work in this paper could be extended in several directions. By considering higher orders of the $1/D$ expansion,
 one may explore other interesting problems. For example, in \cite{Tanabe16050811} the elastic instability is found
 by adding $1/D^2$ corrections to the effective equations and performing the $1/\sqrt{D}$ instead of the $1/D$ expansion. It
 would be interesting to investigate if this occurs for the EGB black ring and what the  end point of the elastic instability
 could be. Moreover, recently it was found \cite{Emparan1802} that by performing the  $1/D$ expansion to higher orders the behavior
 of the black string below the critical dimensions can be clearly captured. It might be possible to study the
 behavior of the black ring below a similar critical dimension.


\section*{Acknowledgments}

B. Chen and P.-C. Li
were supported in part by NSFC Grant No. 11275010, No. 11325522, No. 11335012
and No. 11735001. C.-Y. Zhang is supported by National Postdoctoral
Program for Innovative Talents BX201600005.


\begin{thebibliography}{}

\bibitem{Emparan0801} R. Emparan and H.S. Reall, ``Black holes in higher dimensions," Living. Rev. Rel. {\bf11} (2008) 6
[arXiv:0801.3471].

\bibitem{Horowitz2012} Gary T. Horowitz, {\em Black Holes in Higher Dimensions}. Cambridge University Press, 2012.

\bibitem{Emparan0110260} R. Emparan and H. S. Reall, ``A Rotating black ring solution in five-dimensions,"
Phys. Rev. Lett. {\bf88}, 101101 (2002) [hep-th/0110260].

\bibitem{Dias1510} O.J.C. Dias, J.E. Santos, and B. Way,   ``Numerical Methods for Finding Stationary Gravitational Solutions," Class. Quant. Grav. {\bf33} (2016) no.13, 133001
[arXiv:1510.02804].

\bibitem{Emparan0708} R. Emparan, T. Harmark, V. Niarchos, N. A. Obers and M. J. Rodriguez, ``The Phase
                       Structure of Higher-Dimensional Black Rings and Black Holes," JHEP {\bf0710} (2007)110 [arXiv:0708.2181].

\bibitem{Armas1402} J. Armas and T. Harmark, ``Black holes and biophysical (mem)-branes," Phys. Rev. D {\bf90}
(2014) 124022 [arXiv:1402.6330]

\bibitem{Hovdebo0601079} J. L. Hovdebo and R. C. Myers, ``Black rings, boosted strings and Gregory-Laflamme,"
Phys. Rev. D {\bf73}, 084013 (2006) [hep-th/0601079].

\bibitem{Elvang0608076} H. Elvang, R. Emparan and A. Virmani, ``Dynamics and stability of black rings," JHEP
{\bf0612}, 074 (2006) [hep-th/0608076].

\bibitem{Gregory9301052} R. Gregory and R. Laflamme, ``Black strings and p-branes are unstable," Phys. Rev. Lett. {\bf70}
(1993) 2837 [hep-th/9301052].

\bibitem{Harmark0701002} T. Harmark, V. Niarchos and N. A. Obers,``Instabilities of Black Strings and Branes," Class.
Quant. Grav. {\bf24} (2007) R1-R90, [hep-th/0701002].



\bibitem{Figueras1107} P. Figueras, K. Murata and H. S. Reall, ``Black hole instabilities and local Penrose
inequalities,” Class. Quant. Grav. {\bf28}, 225030 (2011) [arXiv:1107.5785].

\bibitem{Santos1503} J. E. Santos and B. Way, ``The Black Ring is Unstable," Phys. Rev. Lett. {\bf114}, 221101
(2015) [arXiv:1503.00721].

\bibitem{Figueras1512} P. Figueras, M. Kunesch and S. Tunyasuvunakool, ``End point of black ring instabilities and
the weak cosmic censorship conjecture," Phys. Rev. Lett. {\bf116} (2016) 071102
[arXiv:1512.04532]

\bibitem{Emparan1302} R. Emparan, R. Suzuki and K. Tanabe, ``The large $D$ limit of General Relativity," JHEP \textbf{1306}, 009 (2013) [arXiv:1302.6382].

\bibitem{Emparan1406} R. Emparan, R. Suzuki and K. Tanabe, ``Decoupling and non-decoupling dynamics of large
$D$ black holes," JHEP {\bf1407}, 113 (2014) [arXiv:1406.1258].

\bibitem{Emparan1504} R. Emparan, T. Shiromizu, R. Suzuki, K. Tanabe, and T. Tanaka, ``Effective theory of Black
Holes in the $1/D$ expansion," JHEP {\bf06} (2015) 159, [arXiv:1504.06489].

\bibitem{Bhattacharyya1504} S. Bhattacharyya, A. De, S. Minwalla, R. Mohan and A. Saha, `` A membrane paradigm at
large $D$,” JHEP {\bf1604}, 076 (2016) [arXiv:1504.06613].

\bibitem{Suzuki1505} R. Suzuki and K. Tanabe, ``Stationary black holes: Large $D$ analysis," JHEP 1509, 193(2015)
[arXiv:1505.01282].

\bibitem{Bhattacharyya1511} S. Bhattacharyya, M. Mandlik, S. Minwalla and S. Thakur, ``A Charged Membrane Paradigm
at Large $D$," JHEP {\bf1604}, 128 (2016) [arXiv:1511.03432].

\bibitem{Dandekar1607} Y. Dandekar, A. De, S. Mazumdar, S. Minwalla and A. Saha, ``The large $D$
black hole Membrane Paradigm at first subleading order," JHEP {\bf1612}, 113 (2016)
[arXiv:1607.06475].

\bibitem{Bhattacharyya1704}  S. Bhattacharyya, P. Biswas, B. Chakrabaty, Y. Dandekar and A. Dinda, ``The large $D$ black
hole dynamics in AdS/dS backgrounds," [arXiv:1704.06076].

\bibitem{Bhattacharyya1805}  S. Bhattacharyya, P. Biswas and Y. Dandekar, ``Black holes in presence of cosmological constant: Second order in $1/D$," [arXiv:1805.00284].

\bibitem{Suzuki1506} R. Suzuki and K. Tanabe, `` Non-uniform black strings and the critical dimension in the $1/D$
expansion," JHEP {\bf10} (2015) 107, [arXiv:1506.01890].

\bibitem{Emparan1506} R. Emparan, R. Suzuki and K. Tanabe, “Evolution and End Point of the Black String Instability: Large $D$ Solution,” Phys. Rev. Lett. {\bf115}, no. 9, 091102 (2015) [arXiv:1506.06772].

\bibitem{Tanabe1510} K. Tanabe, ``Black rings at large $D$," JHEP 1602, 151 (2016) [arXiv:1510.02200].

\bibitem{Tanabe1511} K. Tanabe, ``Instability of de Sitter Reissner-Nordstrom black hole in the $1/D$ expansion,"
Class. Quant. Grav. {\bf33} no. 12, 125016 (2016) [arXiv:1511.06059].

\bibitem{Emparan1602} R. Emparan, K. Izumi, R. Luna, R. Suzuki, and K. Tanabe, ``Hydro-elastic
Complementarity in Black Branes at large $D$," JHEP {\bf06} (2016) 117,
[arXiv:1602.0575].

\bibitem{Sadhu1604} A. Sadhu and V. Suneeta, ``Nonspherically symmetric black string perturbations in the
large dimension limit," Phys. Rev. D{\bf93} (2016), no. 12 124002, [arXiv:1604.0059].



\bibitem{Herzog1605} C. P. Herzog, M. Spillane, and A. Yarom, ``The holographic dual of a Riemann
problem in a large number of dimensions," JHEP {\bf08} (2016) 120, [arXiv:1605.01404].

\bibitem{Tanabe16050811} K. Tanabe, ``Elastic instability of black rings at large $D$," [arXiv:1605.08116].

\bibitem{Tanabe16050885} K. Tanabe, ``Charged rotating black holes at large $D$," [arXiv:1605.08854].

\bibitem{Rozali1607} M. Rozali and A. Vincart-Emard, ``On Brane Instabilities in the Large $D$ Limit,"
JHEP {\bf08} (2016) 166, [arXiv:1607.0174].

\bibitem{Chen1702}  B.~Chen, P.~C.~Li and Z.~z.~Wang,   ``Charged Black Rings at large $D$," JHEP {\bf1704} (2017) 167,  [arXiv:1702.00886].

\bibitem{Rozali1707} M. Rozali, E. Sabag, and A. Yarom, ``Holographic Turbulence in a Large Number of
Dimensions," JHEP {\bf1804} (2018) 065 [arXiv:1707.08973].

\bibitem{Chen1804} B. Chen, P.-C. Li, Yu Tian and C.-Y. Zhang, ``Holographic Turbulence in Einstein-Gauss-Bonnet
Gravity at Large D at Large $D$,"  [arXiv:1804.05182].

\bibitem{Emparan0902} R. Emparan, T. Harmark, V. Niarchos and N. A. Obers, ``World-Volume Effective Theory for
Higher-Dimensional Black Holes," Phys. Rev. Lett. {\bf102} (2009) 191301 [arXiv:0902.0427].

\bibitem{Emparan0910} R. Emparan, T. Harmark, V. Niarchos and N. A. Obers, ``Essentials of Blackfold Dynamics,"
JHEP 1003 (2010) 063 [arXiv:0910.1601].

\bibitem{Zwiebach1985} B. Zwiebach, ``Curvature Squared Terms and String Theories," Phys. Lett. B {\bf156} (1985) 315.

\bibitem{Boulware1985} D. G. Boulware and S. Deser, ``String-Generated Gravity Models," Phys. Rev. Lett. {\bf55} (1985)
2656.

\bibitem{Wheeler1986}J.T. Wheeler, ``Symmetric Solutions to the Gauss-Bonnet Extended Einstein Equations," Nucl.Phys. B \textbf{268} (1986) 737.

\bibitem{Kobayashi2005} T. Kobayashi and T. Tanaka, ``Five-dimensional black strings in Einstein-Gauss-Bonnet gravity," Phys. Rev. D \textbf{71} (2005) 084005 [arXiv:gr-qc/0412139].

\bibitem{Suranyi2009} P. Suranyi, C. Vaz and L. C. R. Wijewardhana, ``The fate of black branes in Einstein-Gauss-Bonnet gravity," Phys. Rev. D \textbf{79} (2009) 124046 [arXiv:0810.0525].

\bibitem{Brihaye2010} Y. Brihaye, T. Delsate and E. Radu, ``Einstein-Gauss-Bonnet black strings," JHEP \textbf{1007} (2010) 022 [arXiv:1004.2164].

\bibitem{Kleihaus0912}  B. Kleihaus, J. Kunz and E. Radu, ``Generalized Weyl solutions in $d=5$ Einstein-Gauss-Bonnet theory: the static black ring,"  JHEP {\bf1002} (2010) 092 [arXiv:0912.1725].

\bibitem{Chen1511} B. Chen, Z.-Y. Fan, P. Li and W. Ye, ``Quasinormal modes of Gauss-Bonnet black holes at large $D$," JHEP \textbf{1601}, 085 (2016) [arXiv:1511.08706].

\bibitem{Chen1703} B. Chen and P.-C. Li, ``Static Gauss-Bonnet black holes at large $D$," JHEP {\bf1705} (2017) 025  [arXiv:1703.06381].

\bibitem{Chen1707} B. Chen, P.-C. Li and C.-Y. Zhang, ``Einstein-Gauss-Bonnet Black Strings at Large $D$," JHEP {\bf 1710}, 123 (2017) [arXiv:1707.09766].



\bibitem{Kim0711} H. C. Kim and R.-G. Cai, ``Slowly Rotating Charged Gauss-Bonnet Black holes in AdS Spaces," Phys. Rev. D {\bf77} (2008) 024045  [arXiv: 0711.0885].

\bibitem{Wald1993} R. M. Wald, ``Black Hole Entropy is Noether Charge," Phys. Rev. D {\bf48} (1993) 3427 [arXiv:gr-qc/9307038].

\bibitem{Emparan1402} R. Emparan, R. Suzuki and K. Tanabe, ``Instability of rotating black holes: large $D$ analysis," JHEP \textbf{1406} (2014) 106 [arXiv:1402.6215].

\bibitem{Sorkin0402216} E. Sorkin, ``Critical dimension in the black string phase transition," Phys. Rev. Lett. {\bf93}, 031601
(2004) [hep-th/0402216].

\bibitem{Hollands0605106}  S. Hollands, A. Ishibashi and R. M. Wald, ``A Higher dimensional stationary rotating black hole must be axisymmetric," Commun. Math. Phys. {\bf271}, 699 (2007) [gr-qc/0605106].
\bibitem{Emparan1802} R. Emparan, R. Luna, M. Martinez, R. Suzuki and K. Tanabe,``Phases and Stability of Non-Uniform Black Strings," [arXiv:1802.08191].
\end{thebibliography}
\end{document}